\documentclass[sigconf]{acmart}
\usepackage{algorithm2e}
\usepackage{algpseudocode}
\usepackage{caption}
\usepackage{subcaption}
\usepackage{tikz}
\AtBeginDocument{%
  }
\copyrightyear{2025}
\acmYear{2025}
\setcopyright{cc}
\setcctype{by}
\acmConference[CHI '25]{CHI Conference on Human Factors in Computing Systems}{April 26-May 1, 2025}{Yokohama, Japan}
\acmBooktitle{CHI Conference on Human Factors in Computing Systems (CHI '25), April 26-May 1, 2025, Yokohama, Japan}\acmDOI{10.1145/3706598.3714015}
\acmISBN{979-8-4007-1394-1/25/04}

\begin{document}

\title{eXplainMR: Generating Real-time Textual and Visual eXplanations to Facilitate UltraSonography Learning in MR}

\author{Jingying Wang}
\email{wangchy@umich.edu}
\affiliation{%
  \institution{University of Michigan}
  \city{Ann Arbor}
  \state{Michigan}
  \country{USA}
}

\author{Jingjing Zhang}
\email{zjjing@umich.edu}
\affiliation{%
  \institution{University of Michigan}
  \city{Ann Arbor}
  \state{Michigan}
  \country{USA}
}

\author{Juana Nicoll Capizzano}
\email{jcapizza@med.umich.edu}
\affiliation{%
  \institution{University of Michigan}
  \city{Ann Arbor}
  \state{Michigan}
  \country{USA}
}

\author{Matthew Sigakis}
\email{msigakis@med.umich.edu}
\affiliation{%
  \institution{University of Michigan}
  \city{Ann Arbor}
  \state{Michigan}
  \country{USA}
}

\author{Xu Wang}
\email{xwanghci@umich.edu}
\affiliation{%
  \institution{University of Michigan}
  \city{Ann Arbor}
  \state{Michigan}
  \country{USA}
}
\authornote{Both senior authors contributed equally to this work.}

\author{Vitaliy Popov}
\email{vipopov@umich.edu}
\affiliation{%
  \institution{University of Michigan}
  \city{Ann Arbor}
  \state{Michigan}
  \country{USA}
}
\authornotemark[1]
\renewcommand{\shortauthors}{Wang et al.}

\begin{abstract}
Mixed-Reality physical task guidance systems have the benefit of providing virtual instructions while enabling learners to interact with the tangible world. However, they are mostly built around single-path tasks and often employ visual cues for motion guidance without explanations on why an action was recommended. In this paper, we introduce eXplainMR, a mixed-reality tutoring system that teaches medical trainees to perform cardiac ultrasound. eXplainMR automatically generates subgoals for obtaining an ultrasound image that contains clinically relevant information, and textual and visual explanations for each recommended move based on the visual difference between the two consecutive subgoals. 
We performed a between-subject experiment (N=16) in one US teaching hospital comparing eXplainMR with a baseline MR system that offers commonly used arrow and shadow guidance. We found that after using eXplainMR, medical trainees demonstrated a better understanding of anatomy and showed more systematic reasoning when deciding on the next moves, which was facilitated by the real-time explanations provided in eXplainMR.

\end{abstract}

\begin{teaserfigure}
  \includegraphics[width=\textwidth]{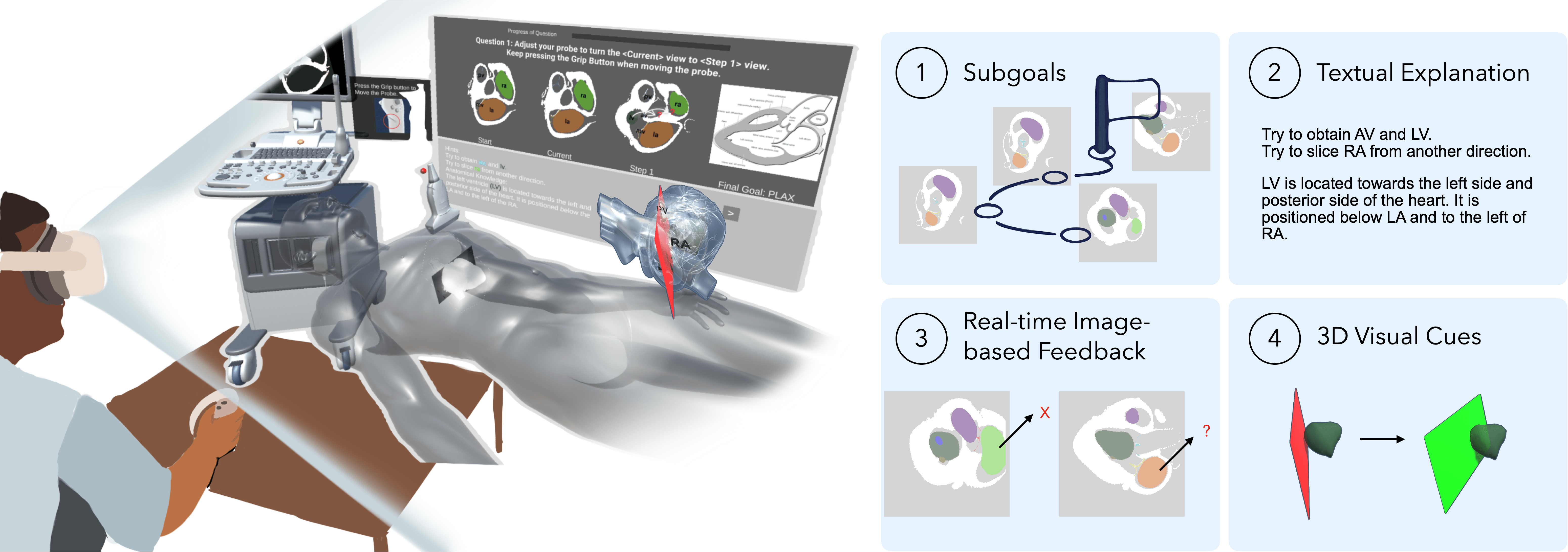}
  \caption{eXplainMR is a Mixed Reality tutoring system designed for basic cardiac surface ultrasound training. Trainees wear a head-mounted display (HMD) and hold a controller, mimicking a real ultrasound probe, while treating a desk surface as the patient's body for low-cost and anywhere training. eXplainMR engages trainees with troubleshooting questions and provides automated feedback through four key mechanisms: 1) subgoals that break down tasks into single-movement steps, 2) textual explanations comparing the current incorrect view with the target view, 3) real-time segmentation and annotation of ultrasound images for direct visualization, and 4) the 3D visual cues provide further explanations on the intersection between the slicing plane and anatomies. }
  \Description{A man sitting in front of a desk wearing a headset to do an ultrasound scan on a virtual patient. }
  \label{fig:teaser}
\end{teaserfigure}

\received{20 February 2007}
\received[revised]{12 March 2009}
\received[accepted]{5 June 2009}

\maketitle

\section{Introduction}
\begin{figure}
    \centering
    \includegraphics[width=0.55\linewidth]{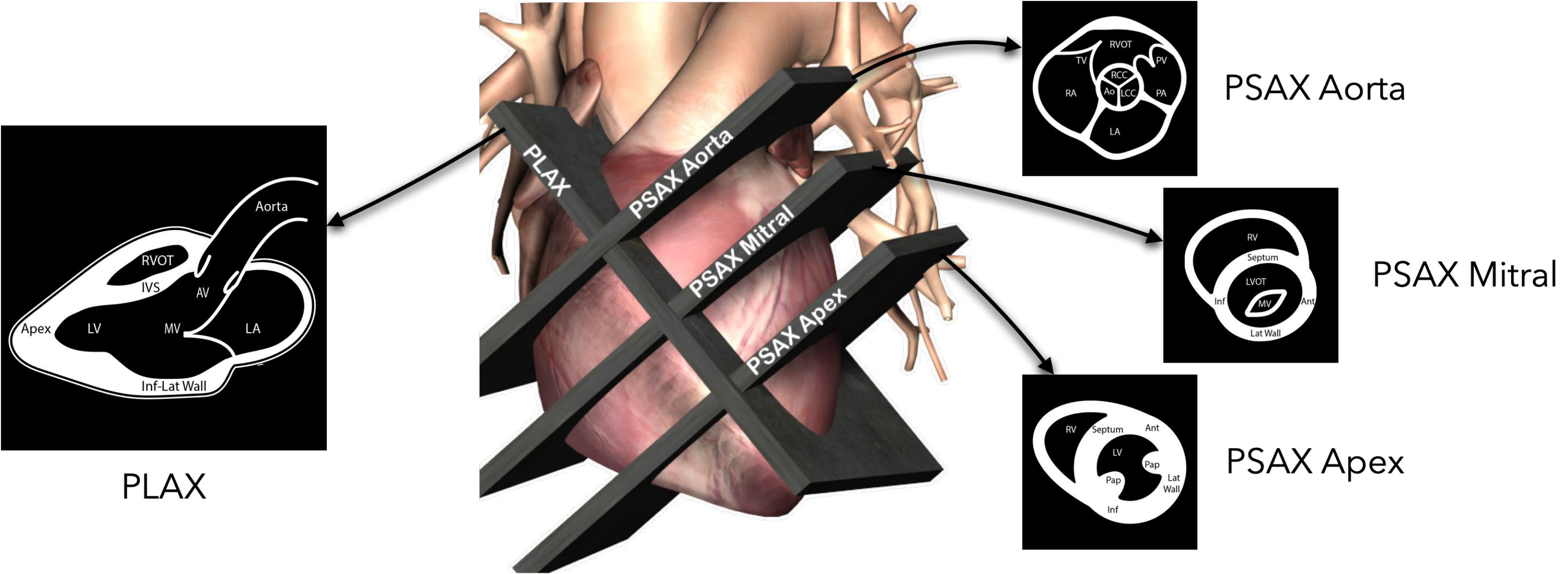}
    \caption{Slicing plane of PLAX and three PSAX view. Obtaining views in PoCUS is similar to getting slices of the heart. }
    \Description{Different slice of the heart.  }
    \label{fig:parasternal}
\end{figure}

Point-of-care ultrasound (PoCUS) has become an essential tool in modern healthcare, improving bedside diagnosis and patient care \cite{dietrich2017point}. PoCUS is a specialized branch of sonography, sharing core principles such as using ultrasound waves to visualize internal structures. Specifically, PoCUS of the heart allows for rapid assessment of cardiac function, detection of life-threatening conditions, and guides critical interventions \cite{CICCONE2004621, devangam2023point}. The core operation of PoCUS was to guide the probe and identify ultrasound views of interest, such as Parasternal Long-Axis View (PLAX) and Parasternal Short-Axis View (PSAX), demonstrated in Fig.\ref{fig:parasternal}\footnote{\url{https://fpnotebook.com/CV/Rad/index.htm}}.
Learning to perform PoCUS requires a trainee to develop a set of visuomotor, visuospatial, and troubleshooting skills \cite{nicholls2014psychomotor}. Visuomotor skills involve forming a mental image of the desired view (examplar) and moving the hands to position the probe for the correct image. Trainees engage in a continuous feedback loop, evaluating whether the obtained view aligns with the visual exemplar and making adjustments until the view meets the exemplar. Visuospatial skills help the sonographer create a 3D mental representation of anatomy based on 2D images. Novice PoCUS learners often struggle to connect theoretical knowledge of ultrasound physics and measurements with the practical application of PoCUS \cite{nicholls2014psychomotor, gibson2020ensuring}. Troubleshooting skills enable trainees to recover from errors and quickly adjust their probe in the process of obtaining images. Although there are instructions on probe placement \cite{ali2020simulator, alty2013practical}, trainees must still learn to adapt when suboptimal views arise due to anatomical variations. This highlights the importance of problem solving skills over the mere memorization of probe positions \cite{boulger2021itsus}.
Developing these skills typically requires hands-on practice under the guidance of an ultrasound expert. The shortage of qualified instructors and high-fidelity simulation equipment poses a significant challenge in training the growing number of learners \cite{situ2021can}. 

High-fidelity simulators \cite{parks2013can, mackay2018can, ali2020simulator, kochan2021point, SilvaJournalOE, SituLaCasse2021CanUN} enhance training by providing 3D visualizations of heart, slicing planes, and correct probe positions, addressing some of the initial challenges (e.g.ultrasound physics and probe placement) faced by novice learners. However, they often overlook troubleshooting suboptimal images and reveal correct answers directly to trainees, focusing on repetitive practice over strategic learning, which limits problem-solving skill development in real world clinical settings \cite{boulger2021itsus}.

Similarly, many intelligent Mixed Reality (MR) tutoring systems \cite{wang2020capturar, huang2021adaptutar, cao2022mobiletutar, chidambaram2022editar, liu2023instrumentar} use virtual instruction to teach a fixed sequence of actions, effective for closed skills (methodically executed each time without variation, e.g., hand washing \cite{nicholls2014psychomotor}). However, PoCUS training involves open skills, requiring adaptable techniques rather than rote-learned fixed sequences to deal with unique anatomical and clinical variables.
Prior MR systems tailored to PoCUS often do not offer explanations for guided instruction \cite{sim1, sim2, sim3, web1, web2, web3phoneinsimo, web4, ar1, vr1, vr2, vr3annotate}, focusing on ``how'' to correct an issue, without addressing ``what'' is wrong and ``why'' specific actions are necessary. 

In this work, we aim to examine how we might automatically generate textual and visual explanations for ultrasound probe movements in an MR environment to enhance schema acquisition \cite{van2017ten} (e.g. rotating the probe to obtain left ventricle in the view, sliding the probe to center the view, etc.) when learners perform cardiac PoCUS, and understand how such explanations influence learning.

To this end, we built eXplainMR, an MR-based cardiac PoCUS training system that automatically generates textual and visual explanations to foster visuomotor, visuospatial, and troubleshooting skills. There are two key features of eXplainMR. First, eXplainMR uses subgoals to scaffold learners and help them practice problem solving in a constrained context. Second, eXplainMR offers real-time explanations to help trainees understand the rationale behind each movement, which is a key difference between our system and existing MR-based training systems. 

The design of eXplainMR is informed by a formative study on expert teaching practices. We found that experts often rely heavily on muscle memory for view adjustments, making it challenging to verbalize their visuospatial and troubleshooting strategies to novices. However, we identified their use of subgoals during path planning: single actions such as moving, rotating, or fanning the probe, often involving familiar or simpler views and highlighting anatomical features relevant to the target view. Furthermore, experts navigate by identifying anatomical structures, comparing them with standard views, and employing mental 3D visualization to guide their probe movements. Based on these findings, we highlight two unique design ideas: 1) Subgoals: use expert strategies to set achievable subgoals involving familiar or simpler views and highlighting important anatomical features. 2) Explanations: provide automated explanations to address the experts' challenge in verbalizing strategies using anatomical comparisons and mental 3D visualization.

To evaluate eXplainMR, we conducted a between-subjects study (n=16) comparing it to a baseline system with conventional single-path, linear guidance (providing commonly used shadow and arrow guidance without explanations), while eXplainMR offered explanations. Both conditions used subgoals. Three experts who teach and practice cardiac PoCUS validated the subgoals as realistic and effective for teaching troubleshooting in cardiac PoCUS. We found trainees in both conditions demonstrated learning gains. In particular, the time it took them to perform a PLAX view significantly dropped from pre- to post-test by 73.1\% on average in eXplainMR, and 55.4\% in baseline. Although the difference was not statistically significant, we gathered rich qualitative data highlighting eXplainMR’s role in helping trainees with probe movements, identifying troubleshooting targets in an incorrect view, mentally visualizing 3D cardiac anatomy, and understanding the correspondence between 3D probe movements and 2D ultrasound images.

The main contributions of our work are:
\begin{enumerate}
    \item A formative study that identified specific challenges in teaching and learning complex visual-motor and visual-spatial skills such as cardiac ultrasound, with a focus on addressing limitations of current single-path task guidance systems.
    \item Development of eXplainMR, a Mixed-Reality (MR) system designed to teach medical trainees how to perform cardiac ultrasound, going beyond simple motion guidance.
    \item An automatic generation pipeline for comprehensive learning scaffolds, including:
        \begin{itemize}
            \item Subgoals for troubleshooting tasks
            \item Textual explanations for recommended moves
            \item Visual explanations based on differences between subgoals
            \item Targeted 3D visualizations to support understanding of cardiac anatomy
        \end{itemize}
    \item A comparative evaluation shows that eXplainMR improves learning in view recognition, 3D mental visualization, and understanding the link between 2D ultrasound and 3D anatomy. It outperforms traditional virtual aids like 3D arrows and shadows. Using subgoals further standardizes hand movements and reduces cognitive load.
\end{enumerate}

\section{Related Works}
\subsection{Mixed-Reality Point-of-Care Ultrasound Training}
 Since the 1970s, ultrasonography training has benefited from the emergence of various high-fidelity simulation technologies \cite{parks2013can, mackay2018can, ali2020simulator, kochan2021point, SilvaJournalOE, SituLaCasse2021CanUN, okano2021outcomes}. These range from mannequin-based devices with sensor-equipped probes \cite{sim1, sim2, sim3, vascularsim, okano2021outcomes} to more advanced solutions utilizing augmented reality (AR) \cite{ar1, evans2024product}, virtual reality (VR) \cite{vr1, vr2, vr3annotate} applications, and even web-based platforms \cite{web1, web2, web3phoneinsimo}. Many of these simulations are now available as commercial products, though often at substantial cost, thereby limiting access in low-resource settings. There is a well-documented body of literature on the effectiveness of simulation-based training in developing competency in basic PoCUS scanning techniques \cite{OlivaresPerez2021VirtualAA, parks2013can, mackay2018can, ali2020simulator, kochan2021point, SilvaJournalOE, SituLaCasse2021CanUN}.

These applications provide various feedback mechanisms, including images and videos demonstrating correct manipulation \cite{vascularsim}, 3D heart model cross-sections \cite{sim1, sim2, sim3, web4}, anatomical structure visualizations \cite{vr1, vr2, vr3annotate, sim3, web4}, guidance for correct probe positioning \cite {sim3,web4}, holograms of ultrasound images over the patient's body \cite{ar1}, and tools for capturing and annotating images \cite{vr3annotate}.

While 3D views enhance anatomical understanding and correct probe positioning provides useful cues, these tools primarily serve as educational aids rather than intelligent tutors.  Without instructors, students may resort to suboptimal strategies like aligning the probe with shadows or memorizing probe positions. Such approaches will fail due to anatomical variability. Instead, trainees must develop the skills to operate the probe by interpreting views, rather than relying on memorized probe positions.

\subsection{Mixed-Reality Tutoring System}

Previous research has demonstrated that Mixed Reality (MR) tutoring offers significant advantages in teaching physical tasks and motor skills \cite{wang2020capturar, huang2021adaptutar, cao2022mobiletutar, chidambaram2022editar, liu2023instrumentar, ipsita2022welding}. MR environments provide immersive guidance, enhancing both the learning experience and the quality of skill acquisition.

Most existing MR tutoring systems focus on areas such as machine or equipment operation \cite{cao2022mobiletutar, chidambaram2021processar, huang2021adaptutar, ipsita2022welding, liu2023instrumentar}, assembly tasks \cite{whitlock2020authar, yamaguchi2020video}, or body coordination activities like physical therapy, exercise, and rehabilitation \cite{anderson2013youmove, faridan2023chameleoncontrol, monteiro2023teachable, semeraro2022visualizing}. These tasks generally fall into the category of closed skills, where learners are expected to replicate or memorize predefined sequences of actions \cite{cao2022mobiletutar, chidambaram2021processar, huang2021adaptutar, ipsita2022welding, liu2023instrumentar}, object placements \cite{ipsita2022welding, whitlock2020authar, yamaguchi2020video}, operational procedures \cite{huang2021adaptutar, ipsita2022welding}, and correct poses \cite{anderson2013youmove, semeraro2022visualizing}. As a result, these systems primarily guide users toward performing predefined ``correct'' actions, often employing instructional tools such as text \cite{whitlock2020authar}, images \cite{anderson2013youmove, cao2020exploratory, chidambaram2021processar}, videos \cite{ipsita2022welding}, 3D arrows \cite{cao2020exploratory, liu2023instrumentar, chidambaram2021processar}, slide bars \cite{liu2023instrumentar, ipsita2022welding}, and avatars \cite{cao2020exploratory, huang2021adaptutar}. In physical therapy and exercise scenarios, visual cues are frequently used to alert users about incorrect movements \cite{anderson2013youmove, faridan2023chameleoncontrol, monteiro2023teachable, semeraro2022visualizing}. 

In contrast, open tasks, like PoCUS, pose unique challenges as learners cannot simply replicate an expert’s solution due to multiple viable approaches and troubleshooting strategies. Systems like Adaptutor \cite{huang2021adaptutar} adaptive hints. However, simply teaching the ``correct'' actions is insufficient as the students may still face difficulties in generalizing them to other cases. Effective training should focus on building adaptive problem-solving skills and explaining the rationale behind actions to align with the open-ended nature of the task. Wang et al. \cite{Wang2023MGPAM} explored instructor-led feedback in VR training, allowing for action explanations similar to traditional classes. However, such resources are often unavailable, and verbal feedback alone is limited in conveying visual information.

Another significant limitation of many MR tutoring systems is their reliance on pre-recorded tutorials that are replayed for users \cite{huang2021adaptutar, chidambaram2021processar, liu2023instrumentar, whitlock2020authar}. While this approach works well for closed tasks, it is far less effective for open tasks like PoCUS, where every case can present unique challenges. Building a tutorial bank with hundreds of specific cases would require enormous effort. Hence, a more intelligent and adaptive system would be better suited to support learning in open-ended and variable tasks.

\subsection{Hand-Eye Coordination}
Hand-eye coordination is typically defined as the ability to control hand movements using visual feedback. During movement, the eyes guide attention toward stimuli, helping the brain understand the spatial position of the body \cite{ballard1992hand, zhu2020hand}.

Research on improving hand-eye coordination includes tasks like moving a cursor to a target using a mouse \cite{smith2000hand, huang2012user}, with added complexity from controllers involving buttons and knobs \cite{sailer2005eye}. In medicine, robotic surgeries exemplify how visual feedback helps surgeons guide tools on-screen \cite{gao2017modeling}, with a key insight being that seeing both current and target positions aids spatial orientation \cite{horstmann2005target}. However, in PoCUS, measuring the distance between the current and target views is more complex than in screen-based tasks. Thus, in designing our tutoring system for PoCUS, it is crucial to provide an easier way to measure this distance.

Training hand-eye coordination involves practicing to achieve a specific goal. Lachman \cite{lachman1997learning} describes this as a stable modification in the stimulus-response relationship due to sensory interactions with the environment. Krakauer and Mazzoni \cite{krakauer2011human} showed that human senses can be trained to reduce false predictions. Repeated practice creates an after-effect that preserves the benefits of training \cite{zhou2023method}. Therefore, offering high-quality practice opportunities is essential in designing effective tutoring systems for hand-eye coordination tasks.

\subsection{Subgoal Learning}
Goal-setting theory by Austin and Vancouver \cite{austin1996goal} suggests that learners tend to break complex tasks into manageable parts, known as subgoals. These intermediate milestones within a larger problem-solving process allow individuals to navigate complex tasks more effectively, track progress, and maintain motivation as they work towards their main objective. 

The effectiveness of subgoal learning in enhancing problem-solving performance has been demonstrated across various fields, such as mathematics \cite{atkinson2003aiding, catrambone1998subgoal, margulieux2018varying}, chemistry \cite{margulieux2018varying}, and programming \cite{margulieux2018varying, margulieux2019finding, margulieux2012subgoal, morrison2020curious, morrison2015subgoals}. Numerous studies in education also emphasize the importance of breaking down problems into smaller subgoals \cite{de2009teaching, guzdial1998supporting, hu2013process, koedinger2013using}. Research indicates that learners with constructive subgoals outperform those using expert-labeled ones in tasks like basic programming and app development \cite{margulieux2019finding, morrison2020curious, morrison2015subgoals} and learner sourcing methods can also generate high-quality subgoals \cite{jin2024codetree, choi2022algosolve}.

While subgoals are well studied in high-cognitive tasks, PoCUS integrates cognitive and physical aspects, making it more of a hand-eye coordination task. In PoCUS, subgoals function more as milestones along the path rather than parts of the solution, resembling path-planning tasks. Unlike in mathematics or programming, where subgoal generation often requires human labeling or learner sourcing, PoCUS benefits from a clear starting point and end goal. Thus, we can explore automatic path planning with milestones as a way to break down the task.

Previous research has explored how robotic arms can determine scan ranges, paths, and poses for each step when scanning a lumbar phantom, which relies on the 3D contours of the skin surface \cite{huang2019robotic}. However, the heart is a much more complex structure, and probe movement in PoCUS is driven by image views rather than surface contours, providing an opportunity for further exploration of automatic subgoal generation.

\section{Formative Study}
We conducted an IRB-approved formative study at a large Midwestern academic healthcare system in the USA with five experts to explore their mental models for obtaining ultrasound views, troubleshooting strategies, and teaching methods. The participants’ demographic information is summarized in Table \ref{tab:formative}. The study aimed to address three key questions in the context of cardiac PoCUS training: 1) What techniques are most effective for troubleshooting incorrect ultrasound views? 2) What specific strategies do experts teach their trainees? 3) What factors contribute to the challenges of learning PoCUS?

\subsection{Procedure}
\begin{table}[]
\begin{tabular}{cccc}
\hline
ID  & Gender & Profession          & Department         \\ \hline
FP1 & Female & Fellow              & Emergency Medicine \\
FP2 & Female & Assistant Professor & Family Medicine    \\
FP3 & Female & Fellow              & Family Medicine    \\
FP4 & Female & Fellow              & Family Medicine    \\
FP5 & Female & Fellow              & Family Medicine    \\ \hline
\end{tabular}
\caption{Demographic information of participants in the formative study.}
\label{tab:formative}
\end{table}

\begin{figure}
    \centering
    \includegraphics[width=0.3\linewidth]{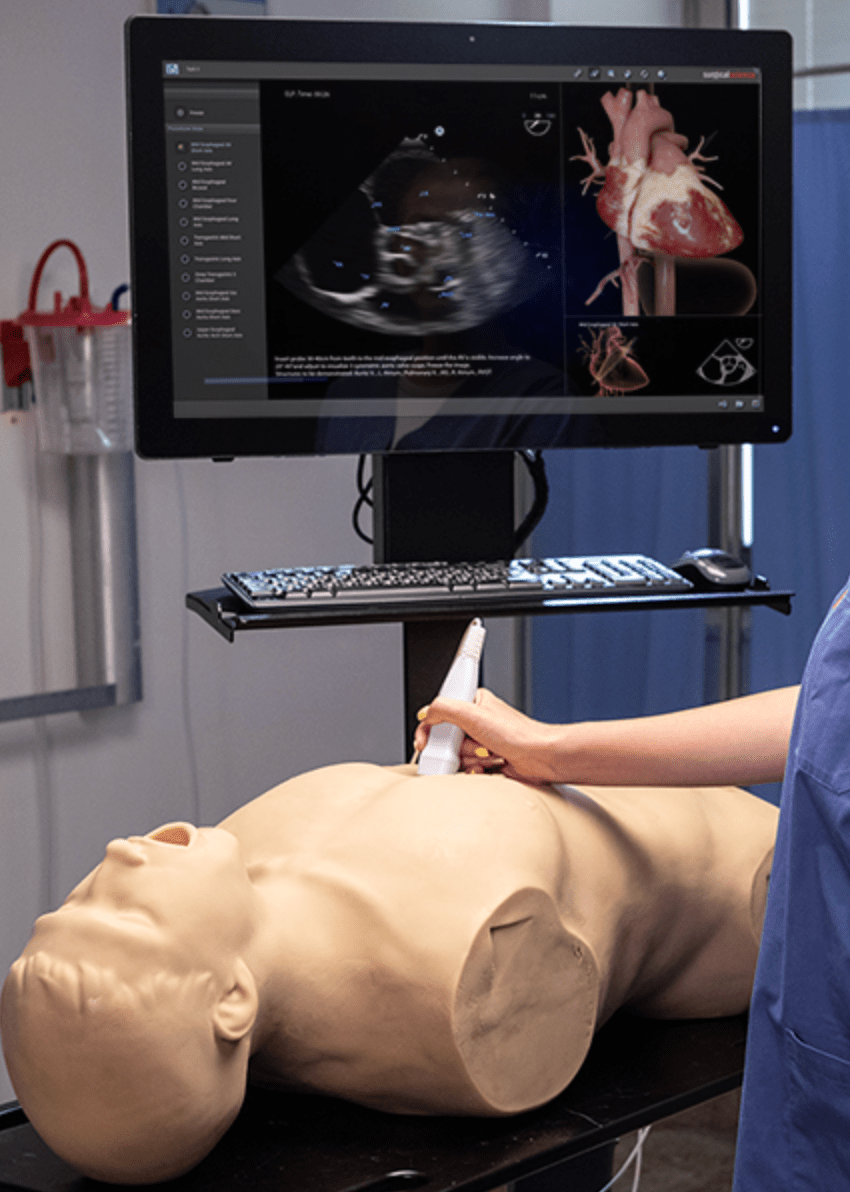}
    \caption{Formative study setup.}
    \Description{A man stands in front of an ultrasound machine, putting the probe on the mannequin.  }
    \label{fig:simulator}
\end{figure}

This study focused on obtaining four key cardiac ultrasound views: the Parasternal Long-Axis (PLAX), Parasternal Short-Axis (PSAX), Apical Four-Chamber (A4C), and Subcostal (SC). Participants engaged in a 1-hour think-aloud session, performing these four views sequentially on the SurgicalScience Ultrasound Mentor \cite{sim3}, a high-fidelity simulator (Fig.\ref{fig:simulator}). They described their step-by-step process for obtaining each view, detailing what they focused on at each stage, the movements they made, and the reasoning behind each movement. The sessions were recorded, transcribed, and analyzed using Affinity Diagram \cite{moggridge2007designing}.
\subsection{Findings}
\subsubsection{The Challenge of Explaining Decision-Making and Troubleshooting in PoCUS Tasks}
Teaching cardiac PoCUS presents unique challenges, as educators often struggle to explain their decision-making and troubleshooting strategies. Instructors commonly point out issues wrong or missing in the current view and suggest corrective actions, or physically guide the trainee’s hand to model specific movements to obtain the optimal imaging plane. However, when asked to explain why certain actions are necessary, many educators attribute it to \textit{``experience''} or \textit{``muscle memory.''} For example, FP1 explained, \textit{``I don’t know, it’s just muscle memory,''} underscoring the instinctive nature of such adjustments.

\subsubsection{Sonographers Use Anatomy Recognition and Comparison to Standard Diagrams for Quality Evaluation}
One of the core skills in cardiac PoCUS is understanding spatial relationships within the heart by interpreting ultrasound images. Educators and advanced learners frequently verbalize their observations, comparing them to standard views of the heart to identify elements missing or incorrect. FP4 said, \textit{``The most important view in this one is that you are going to have all the structures you want to see, like the mitral valve or the papillary muscles.''} FP3 added, \textit{``Instead of just seeing the mitral valve, I need to see a part of the heart, which is the papillary muscle.''} This constant comparison helps learners evaluate image quality and develop the skill of recognizing the view matches the desired standard.

\subsubsection{Both expert and advanced learners employ subgoals when obtaining views.}
It is common for the initial position of the ultrasound probe to be suboptimal, and advanced sonographers often take several steps to adjust the view. Their strategy generally involves setting subgoals with specific characteristics:
\paragraph{The subgoal is usually easy to achieve within one single action.} Sonographers typically make subtle, controlled movements along a single direction, such as fanning or rotating the probe, to locate specific anatomical structures. There are six standard movements in ultrasound: fan, rock, rotate, slide, sweep, and press \cite{alty2013practical}. Instructors frequently use these movements to guide trainees. FP5 explained, \textit{``I basically need to go a little more close to the heart. So I just slide it probably a little bit and then probably rotate to get the marker going exactly to the left shoulder.''}

\paragraph{The subgoal usually contains important anatomical structures that should appear in the target view.}
A key feature of subgoals is the inclusion of critical anatomical landmarks. FP2 noted, \textit{``In this view, I want to see the aorta, I want to see the mitral valve, and all of my adjustments are to get those visualized.''} Another added, \textit{``I am not getting enough of the left ventricle, so I'll probably make adjustments to bring the left ventricle into view.''}

\paragraph{The subgoal can be a familiar or easier view to the trainees.}
Experts often guide trainees through familiar, easier-to-obtain views before advancing to more complex ones. These familiar subgoals create learning milestones, helping trainees troubleshoot more effectively. One participant summarized this by saying, FP1 said \textit{``There are certain movements we’ve memorized as a script: If you see this, do this. If you don’t see this, do that. It’s like there’s an appropriate motion for what you see.''}

\subsubsection{Sonographers relate the 3D anatomy of the heart when driving the probe.}
Experts constantly visualize the 3D anatomy of the heart as they manipulate the probe, relating the acquired 2D images to the heart's structure. FP2 said, \textit{``We kind of try to put the anatomy of the heart in your brain and think about how it should be positioned.''} This mental mapping allows sonographers to better understand how different movements affect the view, as FP3 said \textit{``If I just move a little to the side, I may cut a little too much of the heart, but if I move more to the middle, I can see the right ventricle and mitral valve, which are key structures I need to visualize.''} 

\subsubsection{Sonographers iteratively explore and make adjustments to refine obtained image}
Expert sonographers refine their views through trial and error, experimenting with different movements based on what they observe on the screen. FP1 explained, \textit{``Refining is sometimes based on nothing more than what I’m seeing. If I turn it to the left, does the valve come more into view? If I turn it to the right, does that make it come more into view?''}

\subsection{Design Goals}
Based on the formative study, we identified the following design requirements for developing an intelligent tutoring system to support cardiac PoCUS learning:

\begin{enumerate} 
    \item [D1] Subgoals for Troubleshooting: Break down troubleshooting tasks into manageable subgoals, which should 1) be achievable with a single movement, 2) include key anatomical landmarks in the required view, and 3) relate to easily recognizable or relevant views to guide trainees step-by-step through error correction.
    \item [D2] Key Anatomy Recognition: Enable the recognition and comparison of key anatomical structures to help trainees effectively identify and correct issues in incorrect views. 
    \item [D3] 2D-3D View Integration: Help trainees connect 2D ultrasound views with 3D anatomy, fostering deeper understanding and providing motivation for effective troubleshooting strategies. 
\end{enumerate}

\section{eXplainMR}
eXplainMR is a Mixed-Reality (MR) system designed to help trainees learn cardiac PoCUS. 
eXplainMR provides subgoal-based questions to guide the trainee towards the end goal, emulating troubleshooting scenarios commonly encountered in cardiac PoCUS. The system offers automatically generated subgoals (D1), textual explanations (D2), real-time image-based feedback (D2), and 3D visual cues (D3) to facilitate learning outcomes such as task execution, skill development, and spatial understanding.

\subsection{Interface}
\subsubsection{User Journey}
\begin{figure*}[]
    \begin{subfigure}{\textwidth}
        \includegraphics[width=\textwidth]{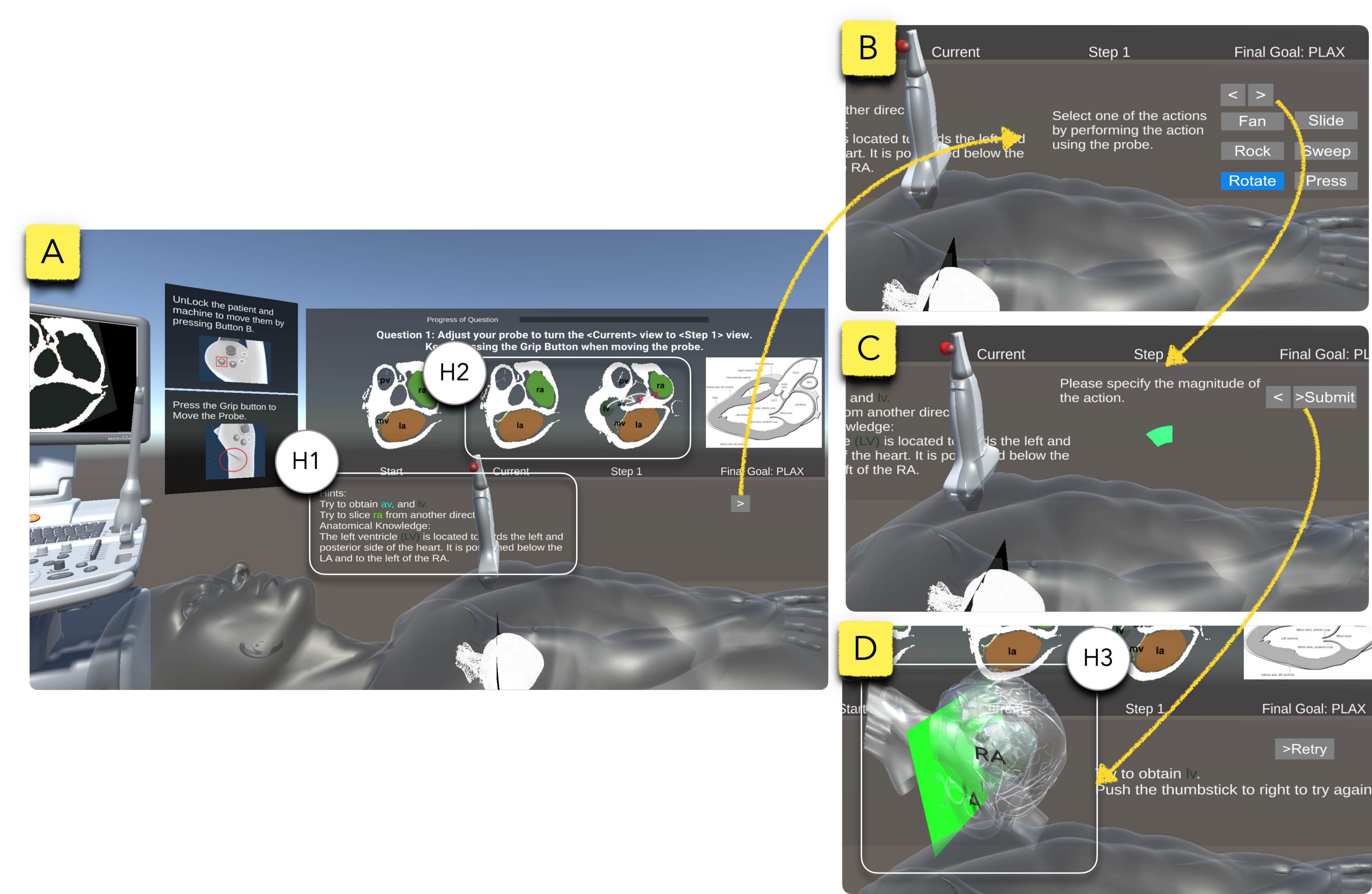}
        \caption{eXplainMR Interface. \fbox{A} shows the free exploration mode, where users receive H1. Textual Explanations and H2. Image-based Feedback, allowing them to try different movements. In \fbox{B}, the user operates the probe, and the system detects the real-time movement made by the user, showing the selected action. In \fbox{C}, the user specifies the amount of movement, and the probe moves steadily in the chosen direction. In \fbox{D}, learners are presented with H3. 3D Visual Cues to further guide their understanding of the movement and anatomy.}
        \Description{4 images of the interface, linked one by one as it means, the panels are triggered in sequence. }
    \end{subfigure}
    \begin{subfigure}{\textwidth}
        \centering
        \includegraphics[width=0.6\textwidth]{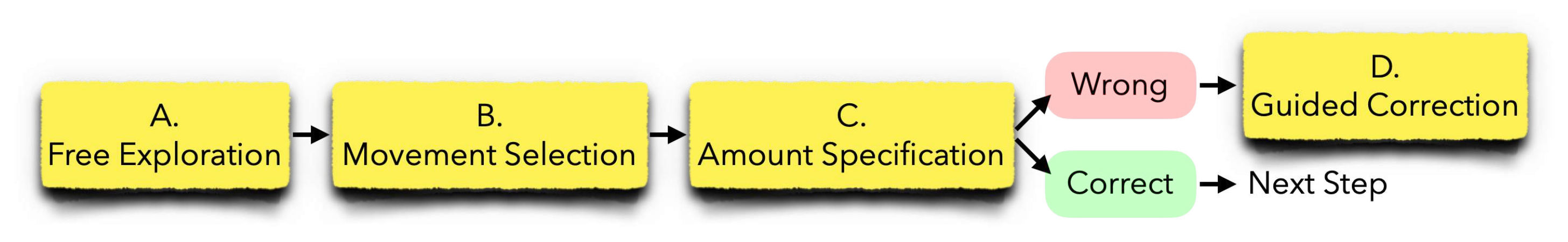}
        
        \caption{eXplainMR Question-Answering Workflow. The user progresses through modes \fbox{A}, \fbox{B}, and \fbox{C} by pushing the thumbstick of the MR controller to the right when they feel ready for the next step. After completing mode \fbox{C}, another push of the thumbstick moves them forward. If an incorrect submission is made, \fbox{D} with 3D visual cues is triggered to assist the user in identifying the error, while a correct submission allows the trainee to proceed to the next stage.}
        \Description{A pipeline diagram describing the user journey. }
    \end{subfigure}
    \caption{eXplainMR QA Platform}
    \label{fig:interface}
\end{figure*}

A medical trainee sits at a desk, and puts on an HMD to enter the eXplainMR training environment. Holding the Pico 4 controller\footnote{\label{pico}\url{https://www.picoxr.com/global/products/pico4}} vertically like a real ultrasound probe, they rest their hand on the desk for stability and adjust the virtual patient to a comfortable, reachable position. Once set up, the trainee begins answering structured questions, with the system initializing the probe at a random starting position. Their task is to move the probe from this initial position toward achieving the PLAX view by completing a series of subgoals. At each subgoal, they follow a structured learning process: (0) they begin at the initial random probe position or the last achieved subgoal; (1) they review textual explanations and image-based feedback detailing the correct probe placement and next steps; (2) they select a movement type from six standard options, such as fanning, rotating, or sliding; (3) they adjust the movement’s magnitude to refine the probe’s position; and (4) they submit their view for evaluation. If incorrect, a 3D visual cue appears, explaining the necessary probe adjustment and prompting another attempt; if correct, they proceed to the next subgoal (see Fig. \ref{fig:interface}).


\subsubsection{Free Exploration Mode}
In this mode, the user is presented with the following: the starting view, the current view obtained by the virtual probe, the target view, and a diagram of the final goal, the PLAX view. The user receives both textual explanations and real-time image-based feedback while manipulating the probe.

The probe is initialized at the starting point, and the user keeps pressing the grip button while moving the probe. Releasing the button freezes the probe at its last position. Users are encouraged to explore the local area around the probe to observe how various probe movements affect the view. Once the trainee decides on a direction, they can push the thumbstick of the controller to the right to proceed.

\subsubsection{Movement Selection Mode}
In this mode, the user must select one of the six standard movements. The system recognizes the selected movement based on the transformation of the virtual probe. After choosing the direction, the trainee can push the thumbstick to the right to proceed.

\subsubsection{Amount Specification Mode}
In this mode, the virtual probe moves only in the selected direction. This provides users with enhanced control, allowing them to read the real-time image-based feedback when checking the alignment between the current view and the target view. The user stops at the optimal view and submits it by pushing the thumbstick to the right.

If the correct view is submitted, the user proceeds to the next step. If the submission is incorrect, the guidance appears as 3D animation visual cues to help users make corrections.

The slicing plane is considered incorrect if its transformation discrepancy from the correct probe position of the target exceeds a distance threshold of 25 units or a rotational threshold of 5 degrees, for a $256\times 256$ slicing plane.

\subsection{Subgoal Generation}
According to D1, achieving the desired PLAX view involves following a step-by-step troubleshooting path. This task can be particularly challenging for beginners, who must manage probe handling and link it to heart anatomy. To aid learners, the task is broken down into subgoals.

\paragraph{Optimization Constrains} Based on findings from our formative study, the subgoals should:
\begin{enumerate} 
    \item Achieve the view with a single movement. 
    \item Bring key structures of the final PLAX view into alignment with the next subgoal. 
    \item Reach a relevant view, such as the parasternal short-axis (PSAX) views, if necessary.
\end{enumerate}

\paragraph{Design Choice} 
The step-by-step optimization method, informed by a formative study, focuses on ``what is the best next step'' rather than treating the sequence of actions as a whole ``what is the best sequence of subgoals''. Predicting entire sequences of subgoals is both cognitively challenging and infeasible due to the variability in ultrasound views. Optimizing single steps is more practical, as experts' step-by-step navigation behavior, identified in the formative study, can be directly searched based on the three constraints. Other methods, such as genetic algorithms \cite{mirjalili2019genetic} and particle swarm optimization \cite{kennedy1995particle}, which rely on fitting functions and search for the entire subgoal sequence, are unsuitable, as they require a definition of ``best sequence of subgoals,'' which is still undefined in the ultrasound imaging context. Therefore, our method mimicking human decision-making by optimizing each subsequent step \cite{becker2022boosting} is more appropriate and efficient.

\paragraph{Searching Process} We model the task of setting up subgoals as a search for views at corresponding positions that meet these constraints, iterating until convergence on the PLAX view. The initial position is randomly sampled. The optimization of subgoals is described in Algorithm.\ref{algo:subgoal}: First, the system samples views along each direction of movement (fanning, rocking, rotating, sliding, sweeping, pressing). Among the sampled images, it attempts to find the view most similar to the target PLAX view to guide the next movement. If the current step is far from the target view and one of the sampled images closely aligns with one of the PSAX views or other familiar views gathered from the formative study, the system will identify this familiar view instead, optimizing the subsequent movement toward the target. The optimization process is executed for each pair of a current view $V_{current}$ and a target view (e.g., PLAX) $V_{PLAX}$ and continues until the current view becomes sufficiently similar to the target view, as determined by a threshold $\epsilon_a$. 

\begin{algorithm}

\KwIn{Target PLAX view $V_{PLAX}$, initial probe position and view $(P_0, V_0)$, movement directions $D = \{\text{fan, rock, rotate, slide, sweep, press}\}$, and familiar views $\{V_{familiar}, P_{familiar}\}$ collected from formative study}
\KwOut{Optimized probe positions $P_0 \rightarrow P_1 \rightarrow \dots \rightarrow P_n$ such that $V_n \approx V_{PLAX}$ at probe position $P_n$}

\SetKwFunction{FindMostSimilarView}{FindMostSimilarView}
\SetKwFunction{SampleViews}{SampleViews}
\SetKwFunction{Similarity}{Similarity}

$P_{current} \leftarrow P_0$ \\
$V_{current} \leftarrow V_0$ \\
\While{\Similarity($V_{current}$, $V_{PLAX}$) < $\epsilon_a$}{
    $Samples = \{(V_i, P_i)\} \leftarrow \SampleViews(D, P_{current})$ \\
    
    $(V_{similar}, P_{similar}) \leftarrow \arg \max_{i} \Similarity(V_i, V_{PLAX})$ \\
    
    \If{
        $\exists (V_{familiar}, P_{familiar}) \in S$ \text{ and }\\
        \Similarity($V_{current}$, $V_{PLAX}$) < \Similarity($V_{familiar}$, $V_{current}$)
    }{
        $V_{current} \leftarrow V_{familiar}$ \\
        $P_{current} \leftarrow P_{familiar}$ \\
    }
    \Else{
        $V_{current} \leftarrow V_{similar}$ \\
        $P_{current} \leftarrow P_{similar}$ \\
    }
    $P_0 \rightarrow P_1 \rightarrow \dots \rightarrow P_{current}$
}
\Return $P_0 \rightarrow P_1 \rightarrow \dots \rightarrow P_n$ \\

\caption{Ultrasound Probe Movement Optimization}
\label{algo:subgoal}
\end{algorithm}

\paragraph{Similarity Measurement} The similarity is measured according to the segmentation map of the view, modeled as the difference between the composition of each structure and the intersection over union (IoU) of the segmentation map. Let $S$ and $L$ denote the segmentation maps of the sampled and target standardized views (e.g., PLAX), respectively, both represented as $n \times w \times h$ matrices, where $n$ is the number of categories, and $w \times h$ is the resolution. $S_i$ and $L_i$ are binary $w \times h$ matrices, with 1 indicating pixels belonging to the $i^{\text{th}}$ category.
: 
\begin{equation} 
    \text{SizeSim}(S, L) = \sum_{i = 0}^{n} 1-\cfrac{||S_i|-|L_i||}{\text{Max}(|S_i|, |L_i|)}
\end{equation}
\begin{equation} 
    \text{PosSim}(S, L) = \sum_{i = 0}^{n} \text{IoU}(S_i, L_i) = \sum_{i = 0}^{n} \cfrac{|S_i \land L_i|}{|S_i \lor L_i|}
\end{equation}
\begin{equation} 
    \text{Similarity}(S, L) = a * \text{SizeSim}(S, L) + b*\text{PosSim}(S, L)
\end{equation}
\begin{figure}[h]
    \centering
    \includegraphics[width=\linewidth]{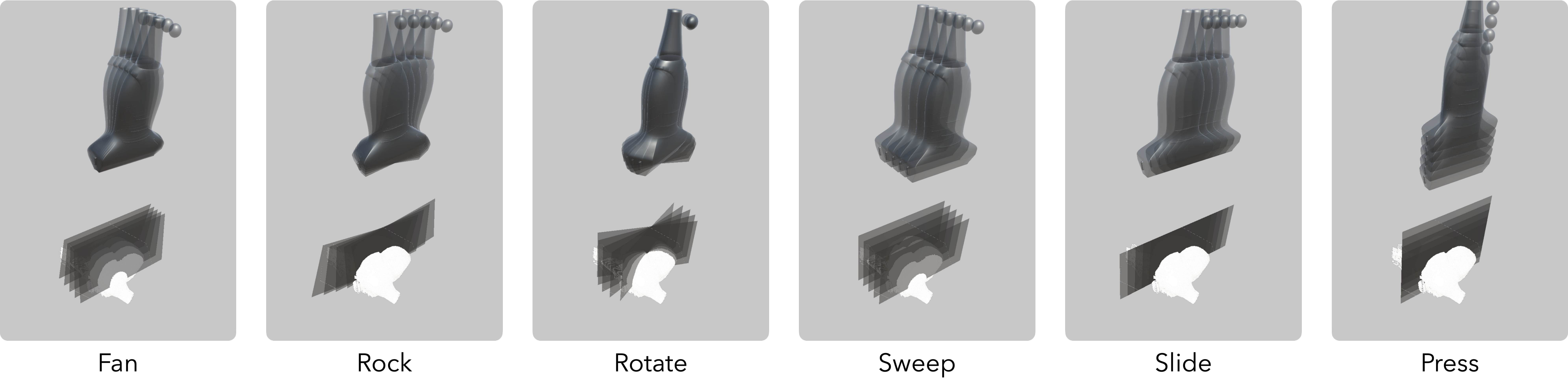}
    \caption{Data sampling process SampleViews$(D, P_{current})$ in Algorithm.\ref{algo:subgoal}. In order to generate the next subgoal, the views along each axis of the 6 movements are sampled.}
    \Description{Plane sliced the heart along 6 movements. }
    \label{fig:datasample}
\end{figure}

\paragraph{Parameter Adaption} The weights for the composition of the structure ($a$), the IoU of the segmentation map ($b$), and the sampling range and number are adapted throughout the subgoal generation process. Before all structures are achieved, $a = 1$ and $b = 0$. The movement search range is $[-1, 1]$ for position and $[-90, 90]$ for rotation, with a sample size of 10 for each direction. After the key structure is achieved, $a = 1$ and $b = 1$. The movement search range is adjusted to $[-0.2, 0.2]$ for position and $[-30, 30]$ for rotation, with a sample size of 20 for each direction. The similarity threshold $\epsilon_a = 10$ is set based on both the composition scores (0–1) for the four chambers and valves, and the IoU scores (0–1) for the segmentation of the four chambers, excluding valves as they are very small in the view images. For each subgoal optimization, we sample 10 or 20 slices, with each sampling taking 1 second to transform the slicing plane and render the slice. On average, 6 steps are required, resulting in a total computation time of approximately 90 seconds for each start-to-end group with subgoals.

\subsection{Feedback \& Hint Generation}
After randomly sampling starting views as troubleshooting questions for PLAX and generating subgoals for each start view, feedback is provided for each subgoal to guide trainees in following the predefined steps. The textual explanations, image-based feedback, and 3D animations are generated based on the comparison of the current and target views, as well as the 3D spatial information of the heart's chambers and valves, thereby providing sufficient 2D and 3D anatomical information. 
\subsubsection{View Segmentation and 3D Anatomy information Extraction}
\begin{figure}
    \centering
    \includegraphics[width=0.5\linewidth]{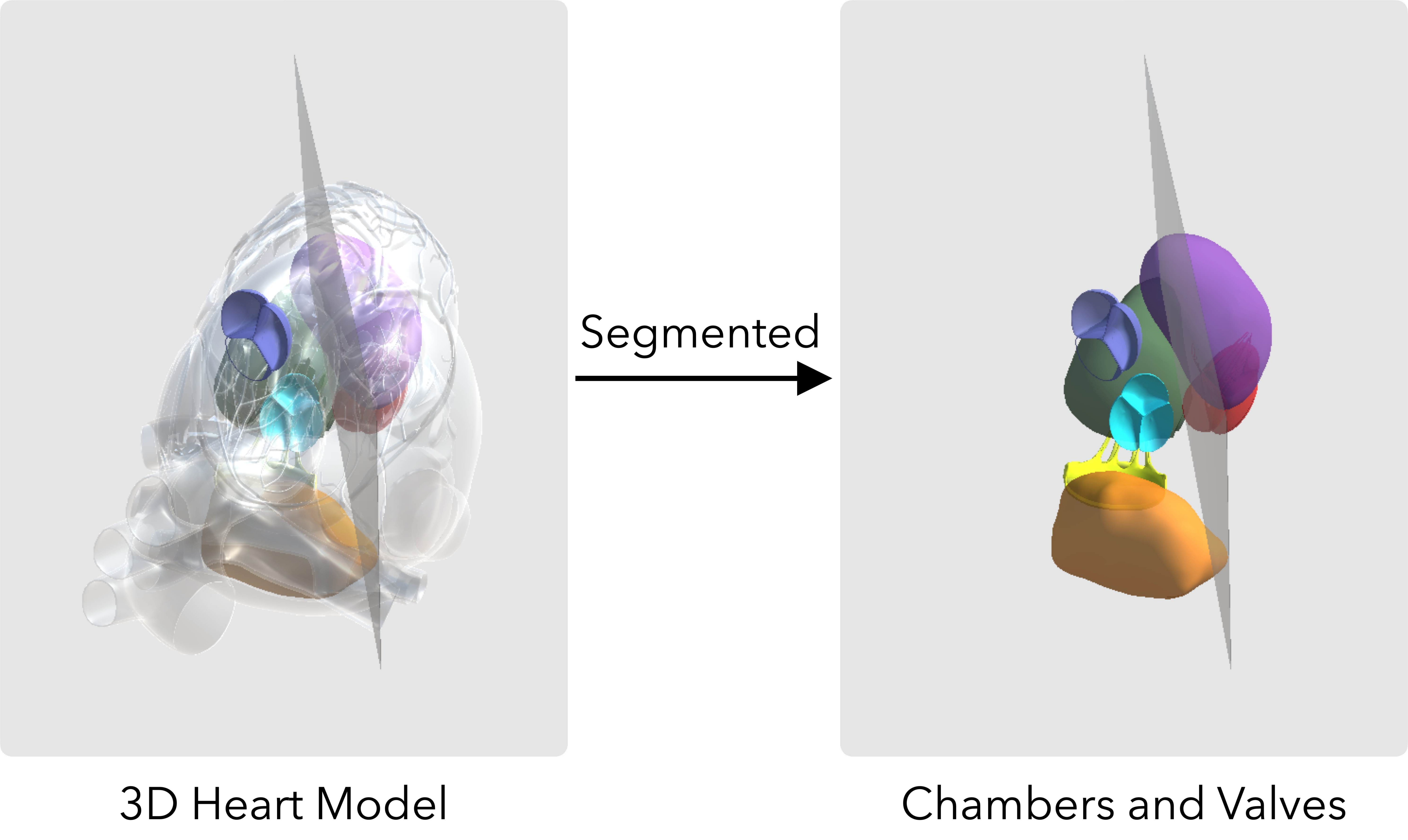}
    \caption{Volume of chambers and valves are extracted from the whole heart model for further view segmentation. The segments include the four chambers: the right and left atria and the right and left ventricles, and four valves: tricuspid valve, pulmonary valve, mitral valve, and aortic valve.}
    \Description{The mask of the heart is peeled off, seeing the chambers and valves.}
    \label{fig:modelparse}
\end{figure}

We generated subgoals and feedback through a two-step process: 1) segmenting the 2D ultrasound view to identify anatomical structures, and 2) extracting corresponding 3D spatial information from the heart model. This enables alignment between visual feedback and anatomical context. The two-step process is accomplished by segmenting the entire 3D heart model. We parsed the 3D heart model in Autodesk MAYA \footnote{\label{maya}\url{https://www.autodesk.com/products/maya}} into volumes representing the chambers and models of the valves, as illustrated in Fig.\ref{fig:modelparse}. We obtain a database storing cross-sections from all directions for each structure (in NRRD format) using 3D Slicer\footnote{\label{slicer}\url{https://www.slicer.org/}}. As the slicing plane intersects the entire heart, it also intersects the individual structures, allowing us to naturally obtain the cross-section of each structure, and thus the segmentation map of the view from these slices. 
\subsubsection{Textual Explanations}
The textual explanation consists of two parts: Problem identification of the start view, and anatomic knowledge of the important structures to help motivate probe movement. It is updated for each subgoal. 

\paragraph{Problem identification of the Start view}
As illustrated in Fig.\ref{fig:texthint}, we compare the structures present in the starting view with those in the target view through the segmentation maps. We identify the structures that are present in the target view but missing in the starting view as missing structures (e.g., the aortic valve [AV] and left ventricle [LV]), and fit them in the template ``Try to obtain [missing structures]''. Conversely, structures present in the starting view but absent in the target view are identified as incorrect structures (e.g., the tricuspid valve [TV]), and fit them in the template ``Try to avoid [incorrect structures]''. Additionally, we compare the shape of structures that exist in both the starting and target views using \texttt{cv2.matchShapes} \footnote{\url{https://docs.opencv.org/3.4/d5/d45/tutorial_py_contours_more_functions.html}}. If the similarity of shape is low, we prompt the user to adjust the slicing plane to obtain a better view of these structures (e.g., the right atrium [RA]: ``Try to slice [RA] from another direction``).

\paragraph{3D Anatomy information}
The 3D anatomical information helps trainees focus on the spatial relationships of key structures. For example, knowing that the left ventricle [LV] is located posteriorly (at the bottom), trainees can orient the probe to target that area (e.g., by fanning to search the bottom). Similarly, understanding that the LV is to the left of the right atrium [RA] enables trainees to adjust the slicing plane of the RA by rotating it to locate the LV. Since multiple structures may be problematic, we prioritize the most important one and provide general positional information relative to the whole heart. 
\begin{enumerate}
    \item If either ``missing'' or ``incorrect'' structures are present, the most important structure is identified as the one with the largest area.
    \item If no ``missing'' or ``incorrect'' structures exist, the most important one is the structure with the largest shape discrepancy.
\end{enumerate} 
In the example shown in Fig.\ref{fig:texthint}, ``Left Ventricle'' (LV) is of the largest area among the missing items, so the information of LV is firstly provided: ``The [LV] is located towards [direction] side of the heart''. 

We also provide contextual 3D positional information for the important structure by describing its relative position in relation to the other two largest structures that are present in both the start and target views. In Fig.\ref{fig:texthint}, the LA and RA are identified as the relevant structures: ``It is positioned [direction] the LA and [direction] RA''. 

The description of direction is generated based on the positional differences \((\Delta X, \Delta Y, \Delta Z)\) between the target structure \((X_0, Y_0, Z_0)\) and either the heart's center \((X_c, Y_c, Z_c)\) or other important structures \((X_1, Y_1, Z_1)\). The directional relationships are defined as follows based on the direction terms in the medical field:
\begin{enumerate}
    \item $\Delta X < 0 $: Left, $\Delta X > 0$: Right,
    \item $\Delta Y < 0 $: Posterior/Below, $\Delta Y > 0$: Anterior/Above,
    \item $\Delta Z < 0 $: Superior, $\Delta Z > 0$: Inferior.
\end{enumerate}

\subsubsection{Real-time Image Feedback}
In addition to rendering the segmented structures in different colors, we label the incorrect segments with ``$\times$'' in the current view and the missing ones with ``$?$'' in the target view, updating these labels in real-time with 60 fps. The identification of the incorrect and missing parts is achieved by comparing the segmentation map of the current view and the target view. In Fig.\ref{fig:imagehint}, at this frame, by comparing ``Current'' and ``Step 1'' view, TV (in red) is incorrectly sliced, and LA and PV are still missing in ``Current'', so TV is labeled by ``$\times$'', and LA and PV are labeled by ``$?$''. 

\subsubsection{3D Animation}
\begin{figure}[]
    \begin{subfigure}[b]{0.23\textwidth}
         \centering
         \includegraphics[height=9em]{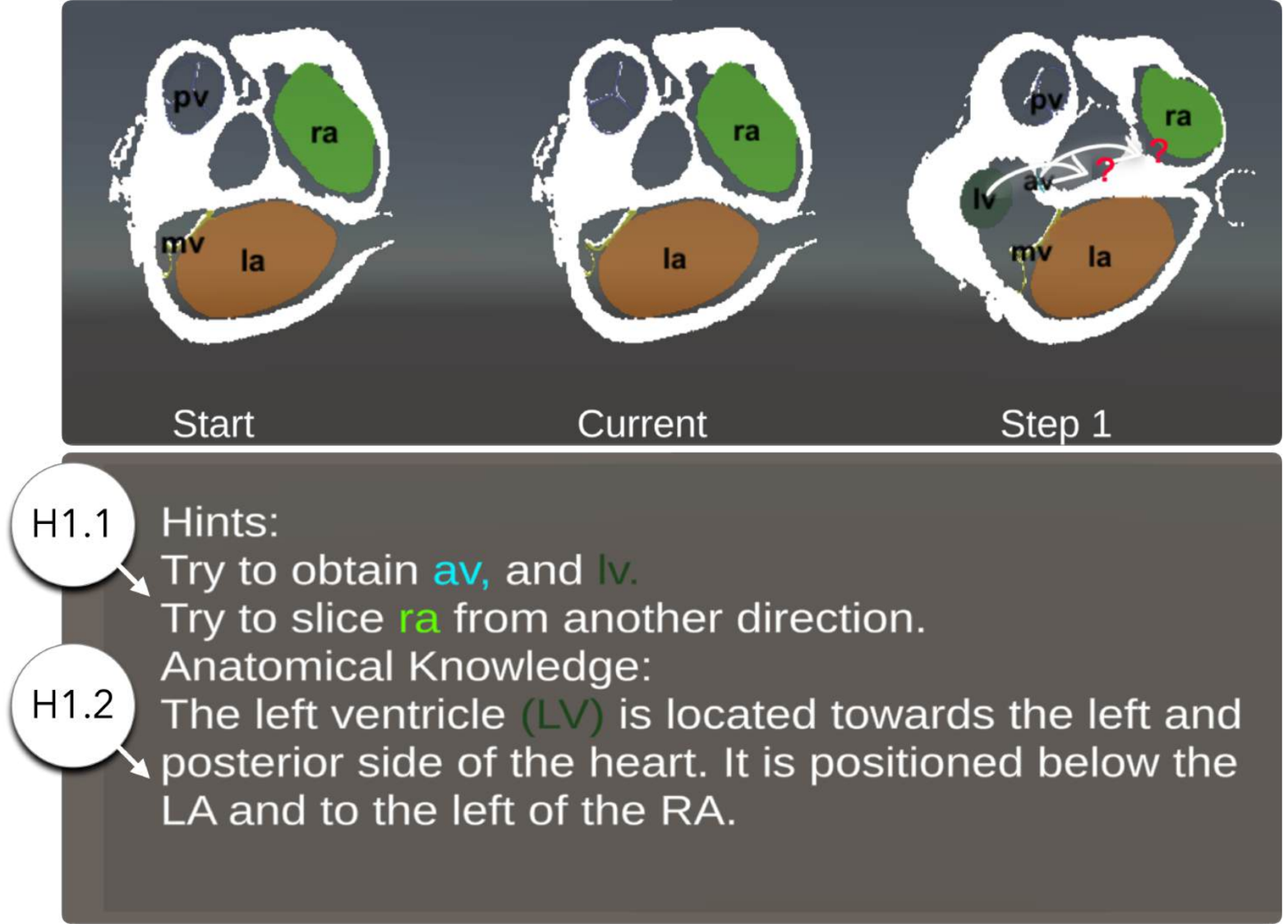}
         \caption{H1. Textual explanation. H1.1 a comparison between the ``Start'' and ``Step 1'' views, and H1.2 automatically identified anatomical details about target chambers or valves.}
         \Description{A virtual textbox showing the content of the textual explanations.  }
         \label{fig:texthint}
         
     \end{subfigure}
     \hfill
     \begin{subfigure}[b]{0.23\textwidth}
         \centering
         \includegraphics[height=6.5em]{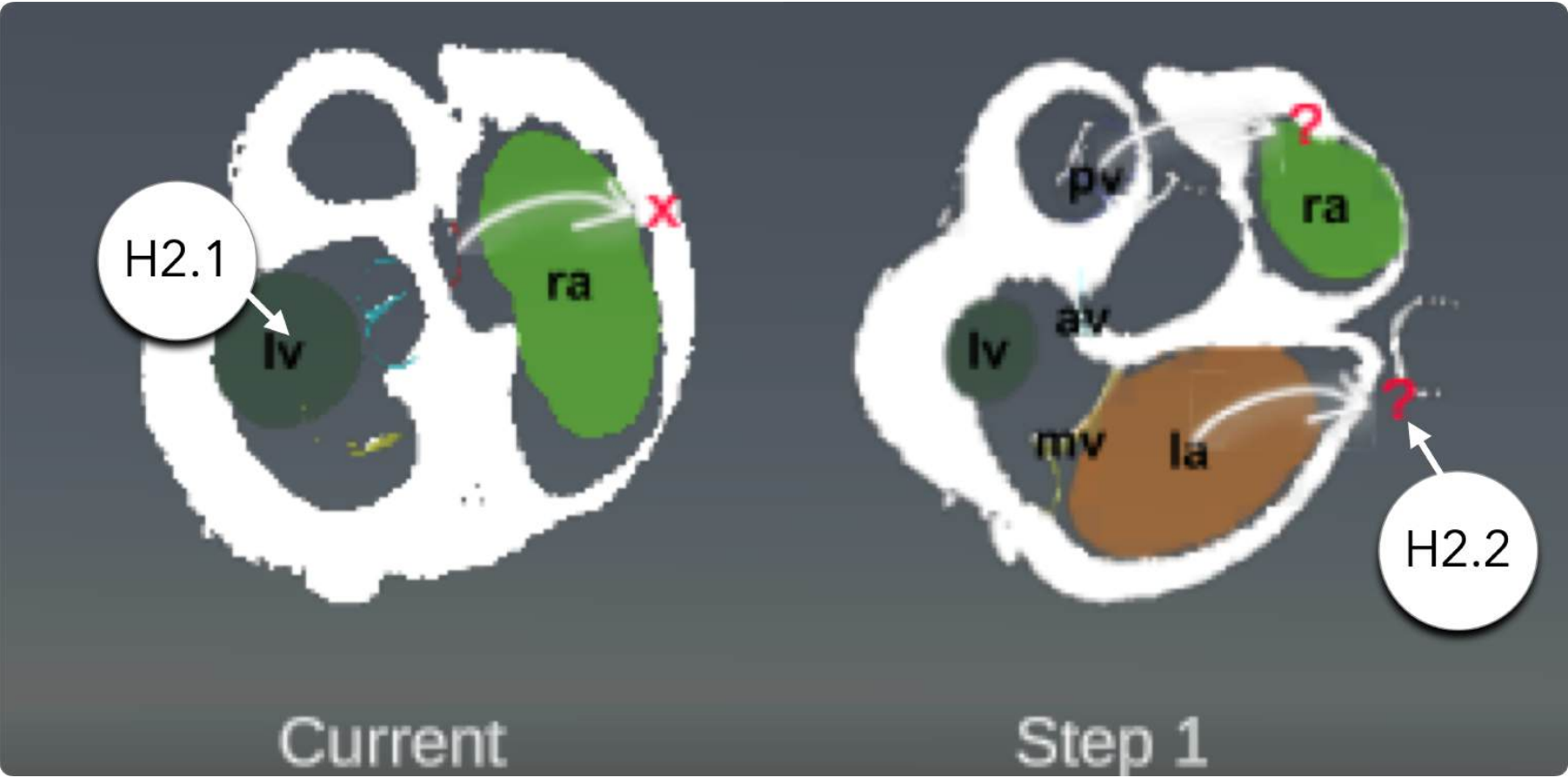}
         \caption{H2. Real-time image based feedback. It consists of two parts: 1) H2.1 real-time rendering of the segments in different colors, and 2) real-time checking of the incorrect and missing structures labeled with $\times$ and $?$.}
         \Description{Two views displayed together and labeled by cross and question marks. }
         \label{fig:imagehint}
     \end{subfigure}

    \begin{subfigure}{0.5\textwidth}
        \centering
        \includegraphics[width=\linewidth]{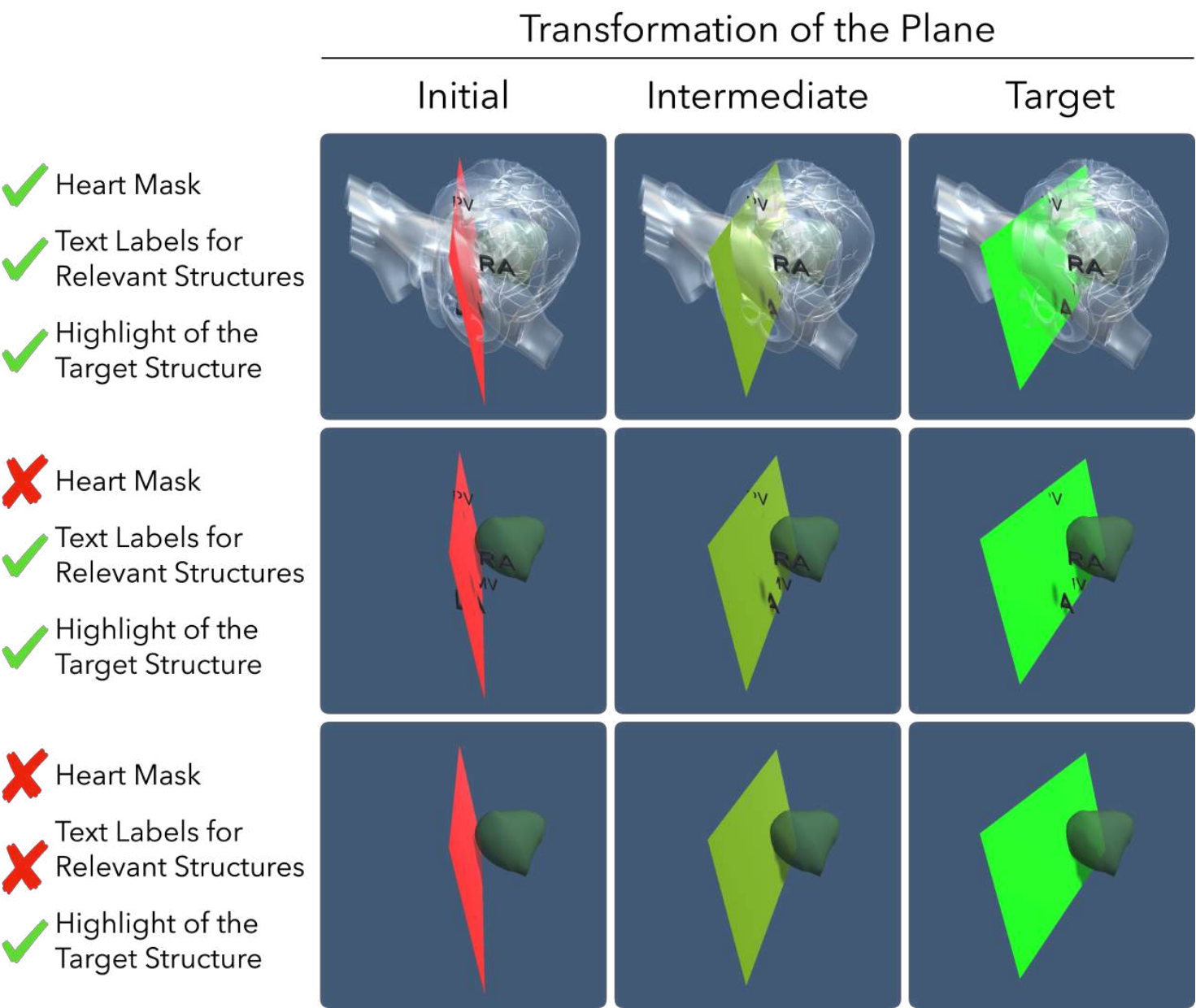}
        \caption{H3. 3D visual cues. The animation shows the plane slicing through the heart (from left to right) using a gradient color scheme, where red represents the starting point and green indicates the target. The visualization is presented in three stages (top to bottom): a full heart view, a key-structure-highlighted view, and a focused display on the most critical target. The animation illustrates how movements will cause the plane to approach, avoid, or slice through the target structure from different angles. }
        \Description{A 3 times 3 table showing the animation from left to right, and level of details increase from top to down. }
        \label{fig:3dhint}
    \end{subfigure}
    \caption{Explantory feedbacks in eXplainMR}
\end{figure}

The 3D animated visual cues include a full visualization of the heart and an animated plane illustrating how the probe should be moved to obtain the correct view. It is updated for each subgoal.  The visualization of the heart’s structure is presented in three levels as shown in Fig.\ref{fig:3dhint}, displayed sequentially:
\begin{enumerate}
    \item Whole Heart Display: Initially, the entire heart is shown.
    \item Semi Focused Display: The heart mask is disabled, highlighting the most important structure. For other structures that appear in the target view, the 3D texts of their names are displayed at their corresponding positions. In Fig.\ref{fig:3dhint}, the model of LV is highlighted and the 3D text of RA, PV, MV, and LA, are also positioned at the corresponding place. 
    \item Focused Display: Finally, only the 3D model of the most important structure is displayed.
\end{enumerate}

This approach provides trainees with a macro-level understanding of how the heart is sliced before gradually transitioning to specific instructions on obtaining or avoiding certain structures, or achieving different cross-sections of a structure.

\subsection{Implementation}
We developed our system on the Pico 4 Enterprise \textsuperscript{\ref{pico}} using Unity3D (2021.2.1f1) \footnote{\url{https://unity.com/cn/releases/editor/whats-new/2021.2.1}}. The user interface was implemented with support from the PICO Unity Integration SDK \footnote{\url{https://developer.picoxr.com/document/unity/}}. To simulate the heart's ultrasound image, we extract the cross-section of the heart. To acquire the image of the cross-section, we first segmented the 3D heart model (MTL format) \footnote{\url{https://www.embodi3d.com/files/category/21-heart/}} manually using Autodesk MAYA \textsuperscript{\ref{maya}}. Then, the segmented model was exported as an NRRD file using 3D Slicer \textsuperscript{\ref{slicer}} and imported into Unity using the UnityVolumeRendering package \footnote{\url{https://github.com/mlavik1/UnityVolumeRendering}}. We iterated on this process with input from experts (see Tab.\ref{tab:formative}) and validated that the level of fidelity achieved in the ultrasound images is sufficient for effective cardiac PoCUS training.
\section{Evaluation on Subgoal Generation}
We evaluated the quality of generated subgoals through an Expert Interview. The study aims to confirm whether our subgoal-generation method reflects an authentic subgoal-generation strategy and whether experts can utilize them for teaching. Our method of generating subgoals is not the only approach, nor is our goal to verify it as the best one. Instead, we aim to demonstrate its feasibility. The learning benefits of subgoals are highlighted in a trainee-focused study presented in Section 7.2.

\subsection{Study Design}
To evaluate the effectiveness of our optimization-based method for generating subgoals, we compared it to a naïve method that breaks down probe movement from start to end in six degrees of freedom. The naïve method serves as a natural baseline for breaking down actions of driving probes with 6 standard movements. The study was designed around a set of predefined troubleshooting scenarios where users would need to navigate from a starting point to the PLAX view.

\subsection{Baseline - Naïve Method}
As the probe transitions from the start position to the end position, a straightforward baseline approach is to decompose the probe's movement into six standard actions, corresponding to its six degrees of freedom (DoF). In 3D space, the probe, as a rigid body, can be represented by its position and orientation as: $P = (\text{PosX}, \text{PosY}, \text{PosZ}, \text{RotX}, \text{RotY}, \text{RotZ}).$ The movement from the start to the end position can then be expressed as the difference vector:
\[
P_{\text{end}} - P_{\text{start}} = (\Delta \text{PosX}, \Delta \text{PosY}, \Delta \text{PosZ}, \Delta \text{RotX}, \Delta \text{RotY}, \Delta \text{RotZ}).
\]
Each movement aligns with an axis or rotation, breaking down probe operation as follows:
\begin{itemize}
    \item \textbf{Fan}: Rotate around X-axis $(0, 0, 0, \Delta \text{RotX}, 0, 0)$  
    \item \textbf{Rock}: Rotate around Y-axis $(0, 0, 0, 0, \Delta \text{RotY}, 0)$  
    \item \textbf{Rotate}: Rotate around Z-axis $(0, 0, 0, 0, 0, \Delta \text{RotZ})$  
    \item \textbf{Slide}: Translate along X-axis $(\Delta \text{PosX}, 0, 0, 0, 0, 0)$  
    \item \textbf{Sweep}: Translate along Y-axis $(0, \Delta \text{PosY}, 0, 0, 0, 0)$  
    \item \textbf{Press}: Translate along Z-axis $(0, 0, \Delta \text{PosZ}, 0, 0, 0)$  
\end{itemize}

\subsection{Study Participants}
We recruited three expert ultrasound educators (see Tab.\ref{tab:expertstudy}) from a large teaching hospital in the United States to participate in the study. All participants have rich experience in teaching and designing PoCUS training curricula.
\begin{table}[]
\begin{tabular}{cccc}
\hline
ID  & Gender & Profession          & Department         \\ \hline
EP1 & Male   & Assistant Professor & Emergency Medicine \\
EP2 & Female & Assistant Professor & Family Medicine    \\
EP1 & Female & Fellow              & Emergency Medicine \\ \hline
\end{tabular}
\caption{Demographic information of participants in the expert study of evaluating subgoals.}
\label{tab:expertstudy}
\end{table}

\subsection{Study Procedure}
The dataset used for evaluation consisted of 10 troubleshooting cases. Each case required learners to navigate from a specific starting position to the PLAX view. For each case, two sets of subgoals were generated: one by the optimization-based method and another by the naïve method. Participants were shown both sets and asked to choose the one they found more intuitive or useful.

\subsubsection{Measurement}
The outcome of the study was based on the experts' preferences for the generated subgoals. We presented the two sets of subgoals to experts, and interviewed them to capture their thoughts on both methods. The interview data was recorded and analyzed using Affinity Diagram \cite{moggridge2007designing} to identify the strengths and weaknesses of each subgoal generation approach, and the inter-rate reliability is reported through Fleiss' Kappa \cite{falotico2015fleiss} score.

\subsection{Findings of Expert Study}
\paragraph{Subgoals generated with the optimization-based method reflect real-world practice by guiding actions based on visual information.}
All experts agreed that the subgoals generated by our optimization-based method closely mirror real-world practices, as they represent meaningful subtasks where trainees must obtain or exclude specific visual structures to align with standard views, rather than mindlessly moving the probe. EP1 remarked when commenting Fig.\ref{fig:subgoal1}, \textit{``The optimization-based method uses a bit of a trial-and-error approach to guide us toward the ultimate goal of ensuring these features are present, indicating that we are on axis.''} EP2 further supported that the subgoals bring important structures, stating, \textit{``That’s realistic—it's very common to first center the left ventricle and then rotate the probe to obtain its long-axis view.''}

\subsubsection{Subgoals generated with the Naïve Method lack educational value, as they contradict the fundamental practice of adjusting views based on visual feedback}.
Experts highlighted that the naïve method is problematic as it generates subgoals based on spatial coordinates, which contradicts the visually driven nature of obtaining views. 
EP1 explained that the Naïve method would fail in teaching, as remembering spatial coordinates is ineffective, \textit{``In reality, the X, Y, and Z axes will differ for every patient. It’s not necessarily a rigid, structured set of movements.''} EP3 echoed this sentiment, confirming that the naïve approach is not practical for teaching novice: \textit{``For novice learners, they’re often not thinking in 3D terms—they’re just randomly moving the probe.''}

\subsubsection{Optimization-based method emulates the process of troubleshooting: first making Large Movements, and then finer adjustments}
EP3 recognized our optimization method because it emulates their process of searching for views: 
\textit{``We make larger movements up and down to see how it looks. Once we find the general movement, then we start making small adjustments, very, very minor movements, to find the view that we're looking for.''}

\subsubsection{Optimization-based method helps recognize important relevant views and teaching common troubleshooting cases}
EP3 commented on the optimization-based method:
\textit{``That's kind of how we teach,''} and further explained, 
\textit{``They just know that if I see this (view), I should move this way. If I see that (view), I should move that way.''} EP3 thinks the subgoals generated with our method are beneficial in building muscle memory for some common troubleshooting cases.  

\subsubsection{Fewer steps are sometimes preferred.}
The efficiency of teaching is also important, as some instructors prefer methods that give fewer steps. In contrast, our method sometimes generates too many steps for adjustments, which can be merged upon request. As EP2 stated:
\textit{``I want to be efficient.''}

\subsubsection{Inter-Rater Agreement}
The expert participants evaluated 10 cases, making binary choices on whether the subgoals generated by our method were better, identical, or worse than the baseline. They agreed in 80\% of cases that our method outperformed the Naïve Method. The inter-rater agreement, measured by Fleiss' Kappa (0.42), reflects moderate agreement, as the predominantly positive votes for our method naturally limit the Kappa score \cite{falotico2015fleiss}. Detailed ratings are shown in Tab.\ref{tab:rate}.

\begin{table}[]
\small
\begin{tabular}{rccccccccccc}
Case ID             & 1 & 2 & 3 & 4 & 5 & 6 & 7 & 8 & 9 & 10 & Sum\\ \hline
Optimization (Ours) & 3 & 1 & 3 & 3 & 1 & 3 & 3 & 3 & 3 & 3 & 26 \\
Naïve (Baseline)  & 0 & 2 & 0 & 0 & 2 & 0 & 0 & 0 & 0 & 0 & 4\\\hline
\end{tabular}
\caption{Expert votes on ``Optimization-Based is better'' vs. ``Naïve Methods is better or equal'' on 10 cases.}
\label{tab:rate}
\end{table}
\section{Evaluation on eXplainMR}
The evaluation study aims to investigate whether the explanations offered in eXplainMR support learning in comparison to frequently used guidance methods in traditional MR tutoring systems that provide step-by-step movement guidance without explanations. We address the following research questions:
\begin{enumerate}
    \item [RQ1] Does eXplainMR support learning of anatomical concepts, and troubleshooting skills of PoCUS?
    \item [RQ2] How do trainees perceive and interact with the feedback (subgoals, text, image, 3D) they receive in eXplainMR?
    \item [RQ3] How do they compare their learning experiences in eXplainMR with baseline condition?
\end{enumerate}

\subsection{Study Design}
To evaluate the effectiveness of eXplainMR for training cardiac ultrasound techniques, we designed a between-subjects experiment where participants completed a series of troubleshooting tasks. In the learning activity, participants in both control and experimental conditions performed PLAX views guided by subgoals, i.e., participants aim to achieve each subgoal which moves them towards the final goal.

\subsection{Baseline}
\begin{figure}
    \centering
    \includegraphics[width=0.5\linewidth]{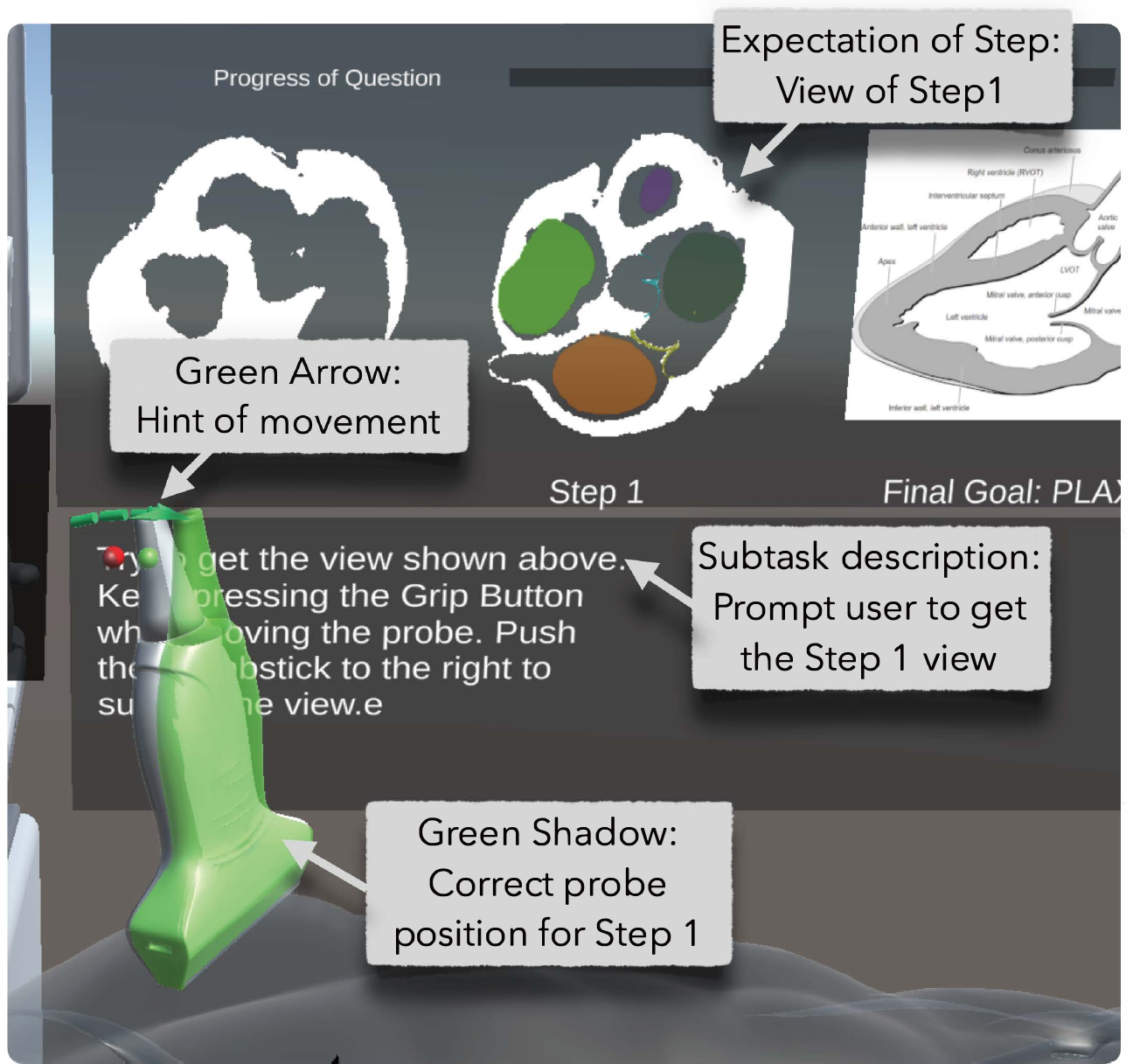}
    \caption{Baseline condition provides green shadow, 3D arrows, subtask description, and step expectation as movement guidance at each step. }
    \Description{The environment for baseline, containing a green arrow and a green shadow. }
    \label{fig:baseline}
\end{figure}

We picked a baseline for control condition that provides commonly used guidance in MR-based tutoring systems \cite{huang2021adaptutar, liu2023instrumentar} for physical tasks, as shown in Fig.\ref{fig:baseline}. Commonly used feedback mechanisms include: (a) an avatar, (b) animated components and arrows, (c) step expectations, and (d) subtask descriptions. In our training scenario, since all movements are centered around the probe, we replace the avatar with a shadow indicating correct probe position. We also use animated arrows to indicate movement direction. For step expectations, specific to PoCUS, the next step's view is displayed, and subtask descriptions are provided through a textbox instructing users to achieve the displayed view saying ``Try to get the view shown above''. 

While both system variants incorporated subgoals, the experimental condition provided textual and visual explanations, allowing us to isolate and evaluate the specific contribution of these explanatory elements to the overall learning experience.
\subsection{Participants}
We recruited 16 participants from a large teaching hospital in the United States, including medical students, residents, and fellows from family medicine, radiology, and emergency medicine. All participants had prior training in PoCUS. Twelve of them completed the entire study procedure, while four of them did not complete a subset of multiple-choice questions in the pre and post-test. We ensured a balanced distribution of participant backgrounds between the two groups. In the eXplainMR group, we included 2 fellows, 5 junior residents, and 1 medical student. In the baseline group, we had 1 fellow, 1 final-year resident, and 6 junior residents. The detailed demographic information is shown in Tab.\ref{tab:userstudy}.

\begin{table*}[]
\small
\begin{tabular}{ccccccc}
\hline
ID  & Gender & Race               & Profession       & Department         &  \# Views & Condition  \\ \hline
P1  & Female & Caucasian          & 1st-yr Resident  & Family Medicine    & 20                              & eXplainMR \\
P2  & Male   & African-American   & 2nd-yr Resident  & Radiology          & 4                               & eXplainMR \\
P3  & Female & Caucasian          & 1st-yr  Resident & Family Medicine    & 50                              & Baseline   \\
P4  & Female & /                  & 1st-yr  Resident & Family Medicine    & 5                               & eXplainMR \\
P5  & Female & Middle Eastern     & 1st-yr  Resident & Family Medicine    & 6                               & Baseline   \\
P6  & Male   & Two or More        & Fellow           & Radiology          & 300                             & eXplainMR \\
P7  & Male   & Caucasian          & Medical Student  & /                  & 3                               & eXplainMR \\
P8  & Female & /                  & 2nd-yr Resident  & Radiology          & 300                             & Baseline   \\
P9  & Female & Latino or Hispanic & 2nd-yr Resident  & Family Medicine    & 10                              & Baseline   \\
P10 & Female & Caucasian          & Fellow           & Emergency Medicine & 250                             & eXplainMR \\
P11 & Female & /                  & 2nd-yr Resident   & Radiology          & 15                              & eXplainMR \\
P12 & Female & Caucasian          & Fellow           & Emergency Medicine & 200                             & Baseline   \\
P13 & Male   & Caucasian          & 2nd-yr Resident  & Radiology          & 300                             & Baseline   \\
P14 & Female & African-American   & 5th-yr resident  & Radiology          & 50                              & Baseline   \\
P15 & Female & Asian              & 2nd-yr Resident  & Family Medicine    & 10                              & eXplainMR \\
P16 & Female & African-American   & 2nd-yr Resident  & Family Medicine    & 10                              & Baseline   \\ \hline

\end{tabular}
\caption{Demographic information of participants in the user study for evaluating eXplainMR. \# Views refers to the estimated number of views performed by the participants. }
\label{tab:userstudy}
\end{table*}
\subsection{Study Procedure}

Each study session lasted approximately 1.5 hours and was conducted in person. 

\paragraph{Pre-test Protocol} Initially, participants completed a pre-test to perform the PLAX view while verbalizing their approach—describing their observations, actions, and reasoning. They also completed a multiple-choice questionnaire (MCQ) \cite{ubcimpocusCardiacQuiz} related to ultrasound anatomy and skills, including 8 minutes to obtain the PLAX view and 12 minutes for the MCQ. The MCQ aimed to assess participants' anatomical knowledge and troubleshooting skills, presenting starting and target views and requiring action selection with justification.
The MCQs covered topics such as: Anatomical Knowledge, Optimized Chamber Size, Avoid/Include Structures, Shifts Between PLAX and Short Axis Views, and Centering the View.

\paragraph{Training Protocol} Participants were then assigned to one of two conditions: the baseline system or eXplainMR. After a 15-minute tutorial to introduce them to the MR system, they spent 20 minutes on the learning activity. 

\paragraph{Post-test Protocol} Upon completion, participants took a post-test with the same format as the pre-test.

\paragraph{After-study Interview} The study concluded with a 15-minute interview to gather qualitative insights on the learning experience using the MR system, and participants completed a demographic survey. The study was approved by the institutional IRB (HUM00256086), and all participants received a \$70 Amazon gift card as compensation.

\subsubsection{Learning Measurement}
Learning outcomes were measured through both the pre-test and post-test. Participants were assessed on their ability to perform the PLAX view, including time taken, action explanations, and MCQ quiz performance. For action-selection questions, participants had to compare their view to the standard and justify their choice of action.

Two expert raters reviewed the responses to assess whether the explanations reflected procedural understanding. Responses such as \textit{``Not 100\% sure''} were rated 0 (invalid), while detailed responses like \textit{``align the aortic root''} or \textit{``capture more of the left ventricle''} were rated 1, signifying strategic understanding.

\subsubsection{Interview Analysis}
The interview recordings were transcribed and analyzed using Affinity Diagram \cite{moggridge2007designing}. Two authors interpreted the transcripts, iteratively grouped the interpretation notes, and identified emerging themes from the data. 

\section{Findings of User Study}
\subsection{Learning Outcomes in Anatomical Concepts and Troubleshooting Skills? (RQ1)}
\subsubsection{Quantitative Outcomes: Both eXplainMR and Baseline improve PLAX view completion time, with eXplainMR showing a weak trend towards greater improvement, whereas there's no significant difference in the MCQ quiz scores.}
We report the learning gains as measured by pre-and post-test performance, including time spent on performing the PLAX view, multiple-choice questionnaire (MCQ) scores, and the quality of responses to open-ended questions.
\begin{figure}
     \centering
     \begin{subfigure}[b]{0.15\textwidth}
         \centering
         \includegraphics[height=10em]{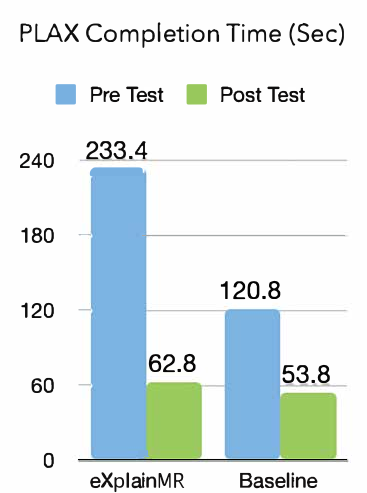}
         \caption{Both eXplainMR and Baseline showed significant improvement on completion time. The eXplainMR group had worse performance before study, but achieved similar level after it.
         }
         \label{fig:plaxscore}
     \end{subfigure}
     \hfill
     \begin{subfigure}[b]{0.15\textwidth}
         \centering
         \includegraphics[height=10em]{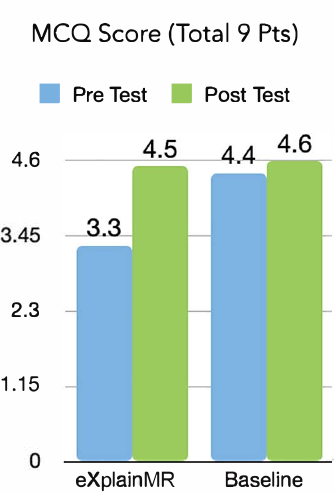}
         \caption{There was modest improvement on MCQ score between the pre- and post-tests. eXplainMR group had lower score before study, but achieved similar level afterwards.}
         \label{fig:mcqscore}
     \end{subfigure}
     \hfill
     \begin{subfigure}[b]{0.15\textwidth}
         \centering
         \includegraphics[height=10em]{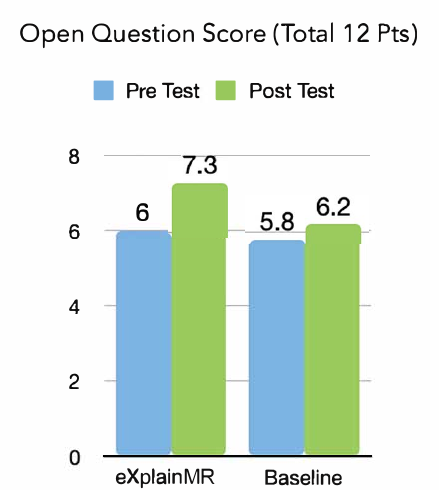}
         \caption{There was a modest improvement in the trainees' quality of explanations of movement choices as demonstrated in the open-ended questions between the pre- and post-tests.}
         \label{fig:openscore}
     \end{subfigure}
     \Description{Histogram of average pre- and post- test performance. }
        \caption{Overall, we observed modest improvement from pre-test to post-test in both conditions, but the difference between the two conditions was not evident. }
        \label{fig:quantscore}
\end{figure}

Among the 8 participants in the eXplainMR condition, 2 were unable to achieve the PLAX view in the pre-test but successfully obtained it in the post-test. The remaining 6 participants, who completed the pre-test, demonstrated a significant improvement in completion time from pre-test (M = 233.4 seconds, SD = 215.01) to post-test (M = 62.8 seconds, SD = 50.36). A paired t-test revealed a statistically significant reduction in completion time, $t(5) = -2.289$, $p = .0419$, indicating that participants completed the PLAX view significantly faster after using eXplainMR. Similarly, among the 8 participants in the baseline condition, 2 were unable to achieve the PLAX view in the pre-test but obtained it in the post-test. For the 6 participants who completed the pre-test, there was also a significant reduction in completion time from pre-test (M = 120.8 seconds, SD = 128.38) to post-test (M = 53.8 seconds, SD = 59.95). A paired t-test showed a statistically significant decrease in time, $t(5) = -2.1092$, $p = .0443$. The normal distribution assumption for t-tests is a weak assumption. It assumes the population data is normally distributed, using the sample distribution as a proxy. For small sample sizes, if the dependent variable is continuous (e.g., time) and assumed to follow a normal distribution in the population, applying t-tests on the sample is valid \cite{SeltmanChapter9}. When comparing the improvement in completion times between the two conditions, participants in the eXplainMR condition showed a larger reduction in time compared to those in the baseline condition. However, the between-group comparison did not reach statistical significance, $t = -1.365$, $p = .1027$, indicating a non-significant trend favoring eXplainMR for speeding up PLAX view completion.

The 6 participants of eXplainMR demonstrated a trend of improvement in their MCQ scores from pre-test (\( M = 3.33 \pm 1.63 \)) to post-test (\( M = 4.5 \pm 1.76 \)) with (\( t(5) = 1.6592 \), \( p = .079 \)). For the 6 participants in the baseline condition, there was a small increase in MCQ scores from pre-test (\( M = 4.4 \pm 0.89 \)) to post-test (\( M = 4.6 \pm 0.89 \)) with (\( t(5) = 0.535 \), \( p = .311 \).

The participants in the eXplainMR condition showed an increase in average scores of open-ended questions from pre-test (\( M = 6.0 \), \( SD = 3.03 \)) to post-test (\( M = 7.33 \), \( SD = 2.07 \)). Similarly, participants in the baseline condition exhibited an increase in average scores from pre-test (\( M = 5.8 \), \( SD = 2.39 \)) to post-test (\( M = 6.2 \), \( SD = 3.11 \)). A comparison between the pre- and post-test scores within each condition reveals an improvement, but there was no statistically significant difference between the eXplainMR and baseline. 

\subsubsection{Participants found eXplainMR to be helpful in movement mastery, contributing to the faster completion of PLAX view.} 
During the think-aloud sessions, participants demonstrated a shift in their approach to performing the PLAX view. In the pre-test, they often made simultaneous movements, reducing precision in probe manipulation. As P1 noted, \textit{``My instinct was always to do multiple movements at once.''} By the post-test, most participants isolated individual movements, making deliberate adjustments with a clearer understanding of their impact on the view. P1 added, \textit{``The whole stepwise process was beneficial... It (eXplainMR) helped me isolate what one movement at a time can do.''} This highlights the effectiveness of subgoals in linking specific actions to view changes. P7 also emphasized the value of certain movements, stating, \textit{``I found the rotation is just so powerful in the MR training, so I tried this movement a lot to see whether I could get the view.''}


\subsubsection{Participants of eXplainMR demonstrated improved problem identification and solution alignment in the post-test.}
Although the statistical improvement in identifying issues in incorrect views and explaining corrective actions was modest, qualitative evidence showed that some participants transferred image comparison skills and 3D anatomical context from eXplainMR.

For example, in the pre-test, P1 frequently relied on vague phrases such as \textit{``Not a complete picture''} and \textit{``I’m not sure''} when comparing views or explaining actions. However, in the post-test, she used specific terms such as \textit{``The left ventricle is not optimally viewed''} and \textit{``I need to move down towards the apex.''} This shift demonstrated a better recognition of the views and a clearer understanding of how her actions influenced the ultrasound view and a deeper grasp of anatomical relationships.
\subsection{Subgoals and Learning Pathways (RQ2)}
\subsubsection{Participants regard subgoals to be beneficial as they offer structured guidance for novices, making it easier to achieve each step.}
Subgoals are particularly beneficial for novice learners who may find it challenging to manipulate the probe and obtain the correct view. As P10 pointed out, subgoals are helpful because without them, \textit{``if I am put in some weird orientation, it would be hard... if you don't know what is the first image I'm trying to acquire? What is the second image I'm trying to acquire?''} By breaking the process into smaller, manageable steps, subgoals provide a structured approach, guiding novice learners and helping them avoid feeling overwhelmed with the changing views or disoriented as they work toward achieving the target PLAX view. P7 echoed this sentiment, stating, \textit{``It's definitely helpful. I feel like good sonography education is going to involve breaking it down into smaller steps because there's just so many variables that are changing all at once.''}

\subsubsection{Participants appreciate that the subgoals allow them to focus on individual movements before mastering the entire procedure.}
Learners can focus on refining specific movements through repeated practice, which helps them understand how each action impacts the ultrasound image. As P13 noted, \textit{``You can focus on one thing at a time and then you don't forget the following steps.''} P14 echoed this by stating, \textit{``It's good for someone to understand if I rotate, this is how it's going to change my image. If I rock, this is how it's going to change my image.''} 

\subsubsection{Subgoals may hinder the ability to develop fluid, real-world ultrasound skills, as it makes participants focus on small transitions too much.}
While subgoals provide valuable guidance, the study highlighted the importance of balancing granularity with fluidity in ultrasound task presentation. Overemphasis on incremental transitions can lead learners to feel that it hinders their ability to connect steps into a cohesive technique. P7 said, \textit{``I found myself just focusing on each individual transition and not the whole process.''} This feedback emphasizes the importance of balancing task decomposition: while breaking down complex movements aids initial understanding, extra chunking may impede procedural fluency.

\subsection{Textual Explanations and Cognitive Overload (RQ2)}
\subsubsection{Anatomical textual explanations help learners imagine heart anatomy, motivate independent problem-solving, and involve participants more in mental visualization.}

The textual explanations that provide anatomical context are valuable for helping learners mentally visualize the 3D anatomy of the heart, as discussed in the formative study. This improved ability of mental imagery supports learners in orienting their hands and encourages them to figure out probe positioning on their own. As P1 explained, \textit{``It helps you put words to the motion you're trying to do. For example, if I was on the right atrium and they wanted me to move to the right ventricle, a reminder that the right ventricle is inferior could make me think, 'Oh right, I need to move this lower,' which would involve this type of motion. I think sometimes it's helpful for putting words to the motion you're trying to do.''} This type of guidance reinforces the connection between anatomical knowledge and probe movement.

\subsubsection{Trainees prefer visual cues over textual comparisons when they are conveying the same message.}
Novice trainees reported that benefited more from visually oriented guidance, particularly labeled images with intuitive symbols like crosses and question marks, than the textual explanations asking them to obtain or avoid structures, as they are conveying the same information. More experienced trainees found detailed textual explanations redundant and overwhelming, preferring subtle visual cues to guide their movements.

\subsection{Image-Based Feedback for Motor-Skill Precision (RQ2)}
\subsubsection{Visual comparison between current and target images enhances eye-hand coordination.}
Direct visual comparison between the current and target ultrasound images allows learners to quickly assess whether their probe movements are bringing them closer to or farther from the desired view, thereby helping them orient themselves effectively. This real-time feedback is crucial for developing hand-eye coordination skills. As P1 noted, \textit{``It was helpful because in real time it was giving me feedback that I’m losing the structures that I wanted''}, while another mentioned, \textit{``I can see two images next to each other and see how they're different.''}

\subsubsection{This segmented view also helps learners recognize anatomical structures and troubleshoot effectively.}
The image-based feedback also helps trainees recognize and interpret anatomical structures in real-time, which enhances their ability to independently troubleshoot and refine their technique.
P4 said \textit{``It gave you some idea what anatomy you are slicing. If they were not labeled for my level of experience, I would've been totally lost.''} and P7 added that \textit{``The labeling on the 2D images was very helpful to understand what each structure is. ''}

\subsection{3D Visual Cues for Movement and Target Orientation (RQ2)}
\subsubsection{Participants found the animated 3D slicing plane helpful in their understanding of how to move toward their goal position.}
The animated 3D visual cues are particularly effective in teaching learners how to transition from their current position to the desired goal, significantly improving their hand-eye coordination—an essential skill for successful ultrasound scans. 
As P10 noted, \textit{``I liked when I got it incorrect, and then it would show me the heart and the actual move I should make, repeating it three or four times. I’d think, okay, that looks like rotating or sweeping. Then I’d look at the image it was guiding me to acquire and ask myself, does it make sense? What other structures would I see if I did this? Then I’d try it, and based on the feedback, it was usually right.''}
\subsubsection{The participants think the highlighted 3D structures provide them with a clear connection between probe movement and target ultrasound views, thereby involving them in the thought process of troubleshooting.}
The 3D model visualization, which highlights key structures, plays a crucial role in helping learners understand the relationship between probe movement and the resulting ultrasound image. This connection is vital for grasping why certain hand motions produce specific views. As P10 mentioned, \textit{``I like focusing on individual (3D) structures. For example, I need to get the left atrium or aortic valve in view. Then I think about what movement is needed to achieve that with its model.''} Similarly, P11 emphasized the importance of understanding anatomical relationships: \textit{``It (3D visual cues) shows you the relationship between the left atrium, left ventricle, aorta, and more, to help you get oriented. That would be more beneficial.''} These reflections underscore the value of breaking down complex heart into smaller, manageable volumes, making it easier for learners to understand how their actions affect the image.
\subsubsection{Participants found eXplainMR's 3D visual cues provide better guidance for troubleshooting and hand-eye coordination than the simulators.
}
eXplainMR's 3D visual cues are specifically designed to guide hand-eye coordination and explain troubleshooting strategies, whereas the simulator \cite{sim3} provides a 3D heart display and cross-sections but does not guide probe movement or explain why the probe should move in a particular way. As P10 noted, \textit{``(3D visual cues on eXplainMR) did a really good job because it tells you how much of each structure was open. When you're dealing with such micro-movements, (I will think), okay, is this valve going to be in view or not going to be in view? I'm definitely getting a left atrium in the view, or I'm definitely not.''} eXplainMR prompts her to consider whether the plane is correctly slicing through the target structure, whereas the simulator simply displays the whole heart, which does not support this type of troubleshooting thought process.

Similarly, when asked how she compares the two kinds of 3D display, P11 commented eXplainMR is better since \textit{``Seeing through the heart and understanding the relationship between the left atrium, ventricle, and aorta helped with orientation''}. This suggests that the Semi Focused Display, the second stage of 3D visual cues (see Fig.\ref{fig:3dhint}) helps learners with hand-eye coordination and motivates probe movement, while the simulator does not have this targeted display.

\subsection{Green shadow and arrows in the baseline are perceived as insufficient for improving learning (RQ3).}
Participants identified a significant limitation in the baseline condition's use of a green shadow for probe alignment. This visual aid inadvertently encouraged learners to focus on the physical probe position rather than the ultrasound screen, contradicting real-world clinical practice where sonographers must continuously monitor the image for adjustments. The consensus among participants was that this feature was counterproductive to learning authentic PoCUS skills. P5 suggested making it optional, stating, \textit{``It's not helpful to have it immediately,''} while P13 expressed concern about its potential to bypass proper technique: \textit{``I felt like I was cheating. I could have didn't even need to look at the screen. I could have just moved to just copy the shadow of where it was supposed to be, which I tried not to do.''} 

The green arrows, some participants think these are still helpful for learning by giving them an idea of what is the correct direction, and they keep focusing on the views, observing the changes, and learning if the view is correct with that movement. P14 said \textit{``But the arrow was nice because it told me, okay, this is the general direction.''}
\subsection{Learning through troubleshooting was found to make eXplainMR a helpful self-learning tool (RQ1).}
According to the participants, the MR system provides a unique question-answering interaction style that facilitates a simpler trial-and-error learning process compared to traditional ultrasound training. P4 highlighted the system's ability to support self-correction: \textit{``I think if I practice it enough, it would be a way to troubleshoot myself over time. Watching myself make a mistake on this is really different than making a mistake on an actual ultrasound, where I don't even know what the mistake is. I know I'm wrong, but I don’t know what I did to mess up the view. So that's why I think it was helpful.''}

\section{Discussion}
Our study on eXplainMR advances MR-based tutoring systems, particularly for tasks requiring high visuospatial cognition and psychomotor skills. Unlike existing systems \cite{chidambaram2021processar, huang2021adaptutar, liu2023instrumentar} that rely on action-based instructions (e.g., arrows, shadows, avatars) for single-path tasks, eXplainMR emphasizes cognitive engagement through two features: (1) subgoal generation to create constrained problem-solving environments, and (2) explanations that provide rationales for each move. This approach surpasses simple motion guidance, fostering deeper learning by encouraging learners to think critically abour each step.
The design of eXplainMR follows ADDIE model, a generic process used in instructional design \cite{gagne2005principles}: Design, Development, Implementation, and Evaluation. Our work contributes to the field through: (1) A formative study that identifies specific challenges in teaching and learning complex visuospatial and psychomotor skills, exemplified by cardiac ultrasound. (2) The development of eXplainMR, an MR system with an automatic generation pipeline that incorporates a set of cognitively stimulating learning scaffolds. (3) A comparative user evaluation that validates the system's efficacy in promoting deeper understanding and skill acquisition. Below we discuss key findings, design implications, address limitations, and outline opportunities for future work.

\subsection{Implications of Providing Explanations for Hand-Eye Coordination Learning}
In hand-eye coordination learning, showing the ``current-target distance'' is highly effective \cite{smith2000hand, huang2012user}, but most ultrasound simulators lack this scaffolding. In contrast, eXplainMR provides real-time image-based feedback and 3D visual cues, indicating both the difference between current and target views and the proximity of the slicing plane to the target structure.

Effect explanations connect superficial skills with the underlying theory. In PoCUS, this means linking probe positioning and slicing through specific structures. By synchronizing 3D heart anatomy with 2D ultrasound images and highlighting key chambers or valves, eXplainMR helps learners visualize how probe movements produce particular views. This aligns with Merrill’s first principles of instruction \cite{merrill2002first}, emphasizing the activation of existing knowledge—3D anatomy—as the foundation for hands-on skills.

Another crucial element is demonstrating the consequences of actions. eXplainMR’s 3D animation shows how particular movements yield the target view, reinforcing the cause-effect link in line with Merrill’s principles \cite{merrill2002first}. This visual guidance, which medical learners often prefer \cite{wang2024surgment}, supports non-intuitive tasks that require trial and error and repeated practice to build muscle memory.

\subsection{Generalization of Automatic Explanation Generation Pipeline}
The potential for generalizing eXplainMR's automatic feedback and explanation generation pipeline extends beyond cardiac PoCUS. Participant feedback indicates the system could be adapted to other ultrasound diagnostics by incorporating 3D models of target organs and anatomy-based segmentation, enhancing its versatility across medical applications.

The principles behind eXplainMR’s action-guidance explanations could also apply to other high-cognitive psychomotor tasks, such as remote drone driving, catheter-based interventions, robotic surgery, and laparoscopic procedures. These tasks share requirements for hand-eye coordination and visuospatial awareness, where learners must perform non-intuitive hand movements guided by visual cues on a screen. MR systems like eXplainMR provide a safe environment for repeated practice, essential for developing the visuomotor, visuospatial, and troubleshooting skills.

\subsection{eXplainMR's Role in PoCUS Education}
eXplainMR is well-suited for beginner to intermediate learners with a basic introduction to cardiac ultrasound. It supports their transition from foundational knowledge to practical application by enhancing hand-eye coordination, image acquisition, decision-making, and interpretation skills. The choice of Mixed Reality (MR) provides key advantages, including natural displays, accurate probe tracking, and a more immersive experience compared to web-based or mouse-controlled simulations. Additionally, MR offers an economical and accessible solution for widespread PoCUS training.

A major benefit of eXplainMR is its self-directed learning environment, enabling learners to practice without the pressure of real patients or direct supervision. This encourages experimentation and skill development at their own pace. By breaking tasks into smaller, achievable steps based on expert heuristics, eXplainMR builds learner confidence before progressing to complex procedures and real-world applications.

The adoption of eXplainMR in medical schools would require planning for equipment costs, training faculty on the eXplainMR, and integrating it into the existing curriculum, (self-) directed practice, and assessment methods. 

\subsection{Limitations and Future Work}
While eXplainMR showed learning benefits, larger-scale studies are needed to validate its impact. PoCUS training takes years, making it hard to assess significant skill gains in a 90-minute session. Multiple-choice assessments help evaluate cognitive skills but do not fully capture hands-on proficiency. Integrating eXplainMR into training curricula could enable long-term evaluation alongside traditional simulations and VR methods \cite{sim1, sim2, sim3, vascularsim, vr1, vr2, vr3annotate, Wang2023MGPAM}. Future research should compare how learners trained with eXplainMR perform versus those taught traditionally and examine how well they can apply these MR-learned skills when scanning real patients.

Using an MR controller instead of a real probe may reduce realism, potentially affecting skill transfer. Advanced sensors could improve fidelity by simulating actual probe handling \cite{yoosaf2022role}. 

Moreover, the current system's predefined optimized paths may not accommodate all learners effectively. No single optimal solution exists for troubleshooting. Learnersourcing \cite{kim2015learnersourcing} could generate diverse subgoal strategies tailored to different expertise levels.

Our study focused on hand-eye coordination for ultrasound acquisition, separate from diagnostic training. Future work could integrate diagnostic skills for a more comprehensive learning experience. eXplainMR’s multi-step interactions may also pose a learning curve for beginners, requiring interface improvements for better usability. user feedback will be essential to enhance usability.

Lastly, incorporating personalized and adaptive feedback mechanisms, such as automated data analysis and real-time performance monitoring, could better address individual learning needs and preferences, thereby improving training outcomes and user experience.

\section{Conclusion}
This study introduces eXplainMR, a Mixed Reality tutoring system designed to enhance the learning of complex visuospatial and psychomotor skills in PoCUS. While traditional MR guidance systems offer the advantage of providing virtual instructions in tangible environments, they often focus on single-path tasks and rely primarily on visual cues for motion guidance without explanatory context. eXplainMR goes beyond simple motion guidance by automatically generating subgoals for obtaining ultrasound images with clinically relevant information. Importantly, the system provides both textual and visual explanations for each recommended move, based on the visual differences between consecutive subgoals. This explanatory feedback fosters deeper cognitive engagement, allowing learners to understand the rationale behind each action in the ultrasound scanning process. By doing so, eXplainMR aims to foster the learning experience from mere instruction-following to active, reasoning-based skill acquisition. The system's approach of combining subgoal generation with explanatory feedback could be adapted to other ultrasound tasks, medical procedures, and complex physical tasks that require strategic decision-making and fine motor skills.

\begin{acks}
This work was supported by the US National Science Foundation (NSF) under grant \#2406218, and the Graduate Medical Education Innovation fund of the University of Michigan Medical School under grant \#U085303. 
\end{acks}
\bibliographystyle{ACM-Reference-Format}
\bibliography{sample-base}
\appendix
\section{Examples of Generated Subgoals}
\begin{figure*}
    \begin{subfigure}{\textwidth}
        \centering
        \includegraphics[height=20em]{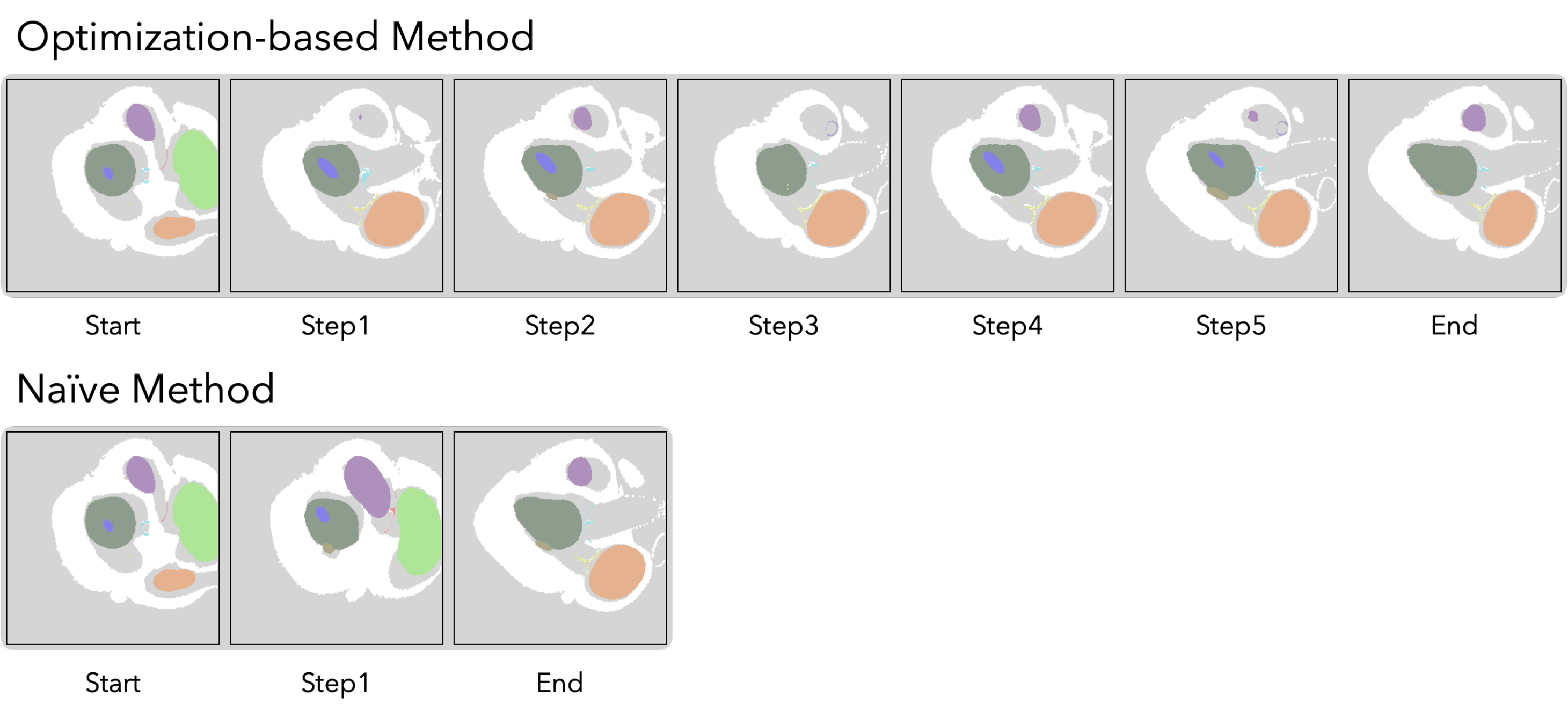}
        \caption{Subgoal Evaluation Case 1: The subgoals generated using the optimization method focused on progressively including the target structures in the PLAX view, and optimizing the shape of the chambers step by step, mimicking the behavior of exploring around to figure out the desired movement. In contrast, the subgoals generated by the naïve method were shorter but more challenging for trainees to follow, as Step 1 did not align with their learning goals or provide a clear path to achieving the desired view.}
        \label{fig:subgoal1}
    \end{subfigure}
    \begin{subfigure}{\textwidth}
        \centering
        \includegraphics[height=20em]{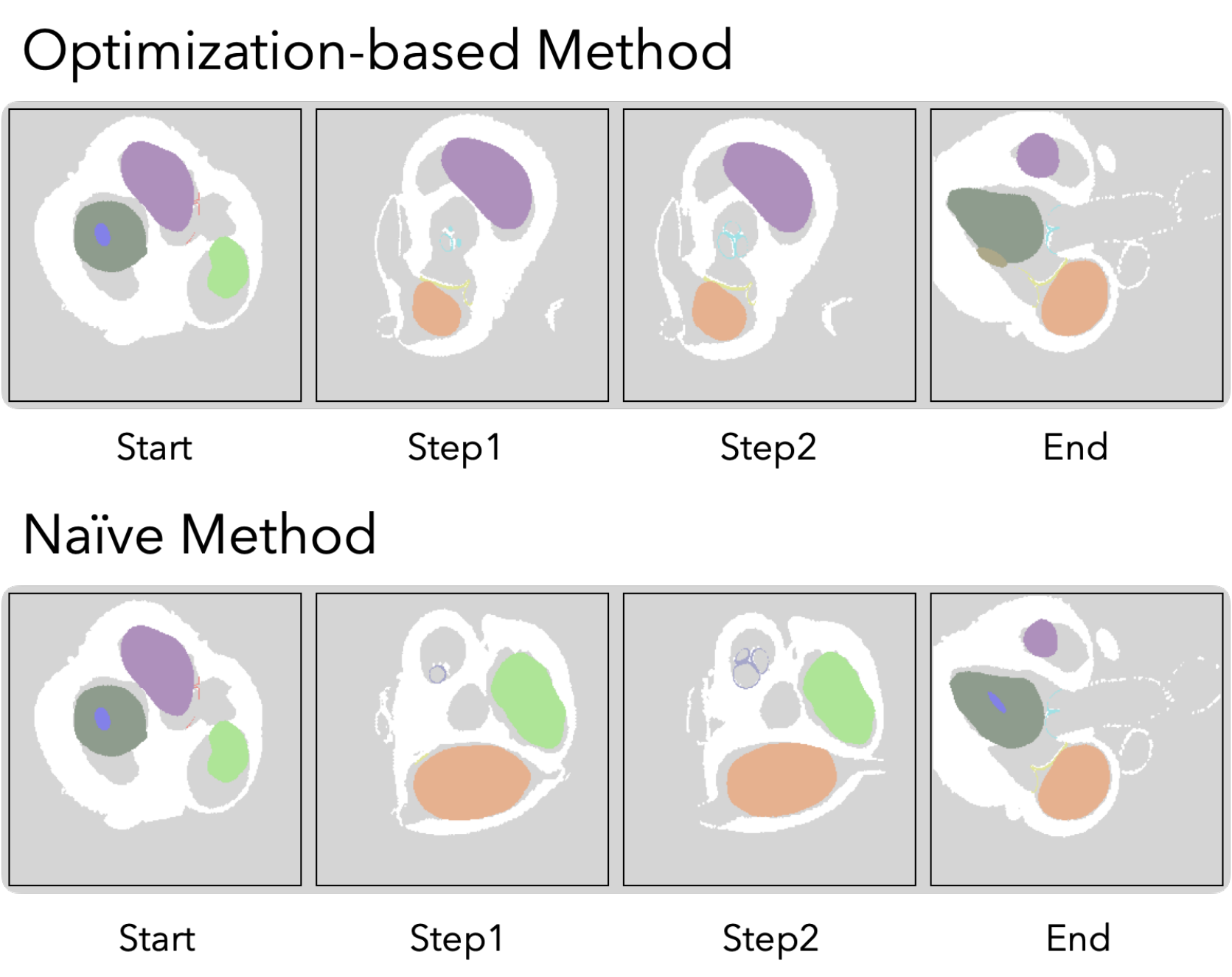}
        \caption{Subgoal Evaluation Case 2: The subgoals generated using the optimization method continued to include the target structures in the PLAX view. The naïve method included the right atrium (green segment), which is not a target structure for the PLAX view, potentially confusing trainees and diverting attention from the primary learning objectives.}
    \end{subfigure}
    \caption{Example of subgoal generation cases. }
    \Description{Sequences of subgoal images. }
\end{figure*}

\end{document}